\documentclass[iop,revtex4]{emulateapj}
\pdfoutput=1
\usepackage[breaklinks,colorlinks,citecolor=blue, bookmarks=false]{hyperref}
\usepackage{pdfpages}
\usepackage[all]{hypcap}
\usepackage{xcolor}
\usepackage{verbatim,epstopdf,graphicx,bigints, amsmath}
\usepackage{array,afterpage}
\usepackage[flushleft]{threeparttable}
\usepackage{pdflscape}
\usepackage{rotating,booktabs}

\newcolumntype{L}[1]{>{\raggedright\hspace{0pt}}m{#1}}
\newcolumntype{C}[1]{>{\centering\hspace{0pt}}m{#1}}

\newcommand{\myemail}{bsalmon@stsci.edu}

\newcommand{\Msol}{\hbox{M$_\odot$}}

\newcommand{\hst}{\textit{HST}}
\newcommand{\jwst}{\textit{JWST}}
\newcommand{\spitzer}{\textit{Spitzer}}

\newcommand{\foot}[1]{\footnotesize #1\normalsize}

\def\Snospace~{\S{}}

\hypersetup{
    colorlinks=true,
    linkcolor=blue,
    filecolor=magenta,      
    urlcolor=cyan,
}
\definecolor{venetianred}{rgb}{0.78, 0.03, 0.08}

\shorttitle{RELICS: The Brightest High-$z$ Galaxies}
\shortauthors{Salmon et al.}

\begin{document}
\title{RELICS: The Reionization Lensing Cluster Survey and the Brightest High-$z$ Galaxies}
\author{Brett Salmon$^{1,\dagger}$, 
Dan Coe$^{1}$, 
Larry Bradley$^{1}$, 
Rychard Bouwens$^{2}$, 
Marusa Brada{\v c}$^{3}$,
Kuang-Han Huang$^{3}$,
Pascal Oesch$^{4}$,
Daniel Stark$^{5}$,
Keren Sharon$^{6}$,
Michele Trenti$^{7}$,
Roberto J. Avila$^{1}$,
Sara Ogaz$^{1}$,
Felipe Andrade-Santos$^{8}$,
Daniela Carrasco$^{7}$,
Catherine Cerny$^{6}$,
William Dawson$^{9}$,
Brenda L. Frye$^{5}$,
Austin Hoag$^{3}$,
Traci Lin Johnson$^{6}$,
Christine Jones$^{8}$,
Daniel Lam$^{2}$,
Lorenzo Lovisari$^{8}$,
Ramesh Mainali$^{5}$,
Matt Past$^{6}$,
Rachel Paterno-Mahler$^{6}$,
Avery Peterson$^{6}$,
Adam G. Riess$^{10, 1}$,
Steven A. Rodney$^{11}$,
Russel E. Ryan$^{1}$,
Irene Sendra-Server$^{12}$,
Louis-Gregory Strolger$^{1}$,
Keiichi Umetsu$^{13}$,
Benedetta Vulcani$^{7}$,
Adi Zitrin$^{14}$
}
\affil{
$^{1}$Space Telescope Science Institute, Baltimore, MD, USA, \\
$^{2}$Leiden Observatory, Leiden University, NL-2300 RA Leiden, The Netherlands, \\
$^{3}$Department of Physics, University of California, Davis, CA 95616, USA, \\
$^{4}$Geneva Observatory, University of Geneva, Ch. des Maillettes 51, 1290 Versoix, Switzerland, \\
$^{5}$Department of Astronomy, Steward Observatory, University of Arizona, 933 North Cherry Avenue, Rm N204, Tucson, AZ, 85721, USA, \\
$^{6}$Department of Astronomy, University of Michigan, 1085 South University Ave, Ann Arbor, MI 48109, USA, \\
$^{7}$School of Physics, University of Melbourne, VIC 3010, Australia, \\
$^{8}$Harvard-Smithsonian Center for Astrophysics, 60 Garden Street, Cambridge, MA 02138, USA, \\
$^{9}$Lawrence Livermore National Laboratory, P.O. Box 808 L- 210, Livermore, CA, 94551, USA, \\
$^{10}$ Department of Physics and Astronomy, The Johns Hopkins University, Baltimore, MD 21218, \\
$^{11}$Department of Physics and Astronomy, University of South Carolina, 712 Main St., Columbia, SC 29208, USA, \\
$^{12}$Infrared Processing and Analysis Center, California Institute of Technology, MS 100-22, Pasadena, CA 91125 \\
$^{13}$Institute of Astronomy and Astrophysics, Academia Sinica, PO Box 23-141, Taipei 10617, Taiwan \\
$^{14}$Physics Department, Ben-Gurion University of the Negev, P.O. Box 653, Beer-Sheva 84105, Israel \\
}
\altaffiltext{}{$\dagger\ $\myemail}
\submitted{Submitted to ApJ}

\keywords{galaxies: high-redshift --- galaxies: evolution --- galaxies: clusters: general --- galaxies: luminosity function, mass function --- gravitational lensing: strong }

\begin{abstract} 
Massive foreground galaxy clusters magnify and distort the light of
objects behind them, permitting a view into both the extremely distant
and intrinsically faint galaxy populations. We present here the
$z\sim 6 - 8$ candidate high-redshift galaxies from the Reionization Lensing Cluster
Survey (RELICS), a \emph{Hubble} and \emph{Spitzer Space Telescope} survey of 41
massive galaxy clusters spanning an area of $\approx$200~arcmin$^{2}$. These clusters were 
selected to be excellent lenses and we find similar high-redshift sample sizes and magnitude 
distributions as CLASH. We discover 321 candidate galaxies with photometric redshifts between $z\sim
6$ to $z\sim 8$, including extremely bright objects with
$H$-band magnitudes of $m_{\rm AB}\approx$ 23 mag.  As a sample, the observed
(lensed) magnitudes of these galaxies are among the brightest known at
$z\geq 6$, comparable to much wider, blank-field surveys. RELICS demonstrates 
the efficiency of using strong gravitational lenses to produce high-redshift samples in the
epoch of reionization. These brightly observed galaxies are excellent targets for follow-up 
study with current and future observatories, including the James Webb Space Telescope.
\end{abstract}

\section{Introduction} 
Images from modern extragalactic surveys are rich with red sources as
we push deeper to reveal the faint, redshifted population of the very
first galaxies. Our investment in this early epoch is for good reason;
the first billion years of the universe ($t_{\rm universe}\approx$ 1~Gyr at
$z=5.5$) cover an era of rapid evolution both in the first stars and
the first galaxies \citep[for a complete review, see][]{Stark16}.
Moreover, this period spans the time when the universe undergoes a
phase transition from being primarily neutral to primarily ionized in
a process called reionization.  Understanding the properties and
relative number of intrinsically faint and bright galaxies at this
epoch directly affects our interpretation of how reionization
occurred, given that the most likely culprits for reionization were
intrinsically faint galaxies at $z>6$ \citep{Madau99, Yan03, Bunker04,
Oesch09, Kuhlen12, Finkelstein12b, McLure13, Schmidt14, Robertson15,
Atek15b, Ishigaki15, Bouwens17b, Livermore17}. 

There have been a variety of approaches to reach this distant galaxy
population. While more costly, deep space-based, blank-field surveys
such as the Cosmic Assembly Deep Extragalactic Legacy Survey
\citep[CANDELS;][]{Grogin11, Koekemoer11} and the Hubble Ultra Deep
Field \citep[HUDF;][]{Beckwith06, Bouwens11a, Ellis13,Koekemoer13,
Illingworth13} as well as wide surveys such as the Brightest of
Reionization Galaxies \citep[BoRG;][]{Trenti11, Bradley12} and
UltraVista \citep{Scoville07,McCracken12,Bowler12,Bowler17} have
produced exquisite datasets that comprise some of the largest and
brightest samples of $3<z<10$ galaxies. The recent
ground-based $z=6-7$ samples from the GOLDRUSH
\citep{Harikane17,Ono17} and SILVERRUSH
\citep{Konno17,Shibuya17a,Shibuya17b,Ouchi17} surveys have discovered 
thousands of high-$z$ galaxies and valuable insight into the behavior of 
Lyman-$\alpha$ 1216~\AA\ emission in the epoch of reionization.  These
surveys continue to surprise us with results from the distant
universe, including the inexplicably bright, most distant confirmed
galaxy found to-date at $z=11.1$ 
\citep[$t_{\rm universe}\approx$ 400~Myr;][]{Oesch16}. 

Another approach is to take advantage of natural telescopes by
observing strong gravitational lenses.  Cluster lensing surveys such
as the Cluster Lensing and Supernovae Survey with Hubble
\citep[CLASH;][]{Zheng12, Postman12, Coe13, Bradley14, Bouwens14} and
the Hubble Frontier Fields \citep[HFF;][]{Coe15, Atek15a, Atek15b,
Lotz17, Ishigaki17, Bouwens17b} have produced most of the $z\geq 8$
galaxy candidates and allowed us to make the first inferences of the
star-formation rate density at $z=9-10$ \citep{Zitrin14, Ishigaki15,
Oesch15}. Moreover, the magnifications produced by lensing enables us to
reach intrinsically faint, low-mass galaxies.  Thanks to carefully
calibrated lensing models \citep[e.g.,][]{Meneghetti17}, subtraction
of intracluster light \citep[][]{Merlin16, Livermore17}, and
calibration of the measured sizes \citep{Kawamata15, Bouwens17a},
there has been substantial progress in deriving both the 
prevalence of intrinsically faint, lower-luminosity galaxies
and the faint-end slope of the ultraviolet (UV) luminosity function (LF)
\citep{Atek14, Atek15b, Alavi16, Livermore17, Bouwens17b}.

In addition, it is important to find highly magnified galaxies
in order to detect intrinsically faint UV metal lines, such as \ion{C}{4}
$\lambda$1548~\AA\ \citep{Stark14, Stark15b} and \ion{C}{3}]
$\lambda$1909~\AA\ \citep{Rigby15, Stark14, Stark15a, Stark17,
Mainali17} at high redshift. These UV lines can now be seen out to $z$=6-7, 
including intrinsically faint Lyman-$\alpha$ \citep{Hoag17}. It is imperative to detect these faint
metal lines not only because they help us to deduce the shape of the
ionizing spectra, but they also allow us to spectroscopically confirm
the redshifts of galaxies in the epoch of reionization, given that the
Lyman-$\alpha$ line becomes completely opaque to the
line-of-site neutral intergalactic medium \citep{Stark10, Schenker12,
Tilvi13, Pentericci14}. 

The rich history of using strong lensing systems to study in detail
$z\approx 4-7$ galaxies \citep{Franx97, Bradley08, Zitrin12, Jones13,
Kawamata15} and reveal the $z\approx 8-11$ population \citep{Zheng12,
Coe13, Bouwens14, Zitrin14, McLeod16, Ishigaki15, Ishigaki17} was the
motivation for the Reionization Lensing Cluster Survey (RELICS; Coe et
al. in prep). RELICS is a 190-orbit \emph{Hubble Space Telescope}
(\hst) Treasury Program designed to build off of the success of other
\hst\ lensing surveys like CLASH and the HFF, and take advantage of
clusters with existing \hst /ACS imaging and/or data suggesting
exceptionally high cluster masses.  In short, the survey targeted 41
massive galaxy clusters selected by the \emph{Planck} survey
\citep{Planck16, Robertson15} to be excellent lensing systems.  This
survey is timely in advance of the \emph{James Webb Space Telescope} 
(\jwst) 2018 launch date, as \jwst\
was not designed to be a wide-field survey telescope and will benefit
from existing high-redshift candidates. We present here the first
results of the RELICS program, providing to the community all of its
high-redshift candidates found to-date.

This paper is organized as follows. In Section~\ref{sec:DataSample},
we summarize our observations, redshifts, and selection. In
Section~\ref{sec:Results} we describe our resulting magnitudes of the
objects in our sample, and present the SEDs and images of bright sources. 
In Sections~\ref{sec:Conclusions} we discuss
our conclusions and future work.  Throughout, we assume concordance
cosmology using $H_0$ = 70 km s$^{-1}$ Mpc$^{-1}$,
$\Omega_{\text{M,0}}$ = 0.3 and $\Omega_{\Lambda,0}$ = 0.7.  All
magnitudes quoted here are measured with respect to the AB system,
$m_{\text{AB}}$ = 31.4 -- 2.5 $\log$($f_{\nu}/1$ nJy) \citep{Oke83}. 

\section{Data, Redshifts, and Sample Selection}\label{sec:DataSample}
\subsection{RELICS Cluster Selection and \hst\ Photometry}\label{sec:relicsData}
The RELICS clusters were selected by a combination of their cluster
mass and pre-existing ACS imaging. From the most massive Planck
clusters \citep[identified by their Sunyaev Zel'dovich cluster
mass;][]{Planck16}, we first selected the 8 most massive clusters that
had \hst/ACS but not WFC3 infrared imaging, and another 13 massive
Planck clusters that had no \hst\ or \spitzer\ imaging at all. The 20
other RELICS clusters are selected from known strong lenses that have
already have \hst\ optical imaging.  We also note that seven of the
RELICS clusters can be found in the MACS program by \cite{Ebeling01}.
We initially inferred the cluster lensing strengths from a variety of
sources, including their X-ray mass
\citep[MCXC;][]{Piffaretti11,Mantz10}, weak lensing mass
\citep{Sereno15, Applegate14, vonderLinden14,Umetsu14, Hoekstra15},
SDSS data \citep{Wong13, Wen12}, and other SZ mass estimates
\citep{Bleem15, Hasselfield13}.  Further details on the cluster
selection can be found by \cite{Cerny17} and Coe et al. (in prep). 

\begin{figure} \epsscale{1.1}       
\centerline{\includegraphics[scale=0.52]{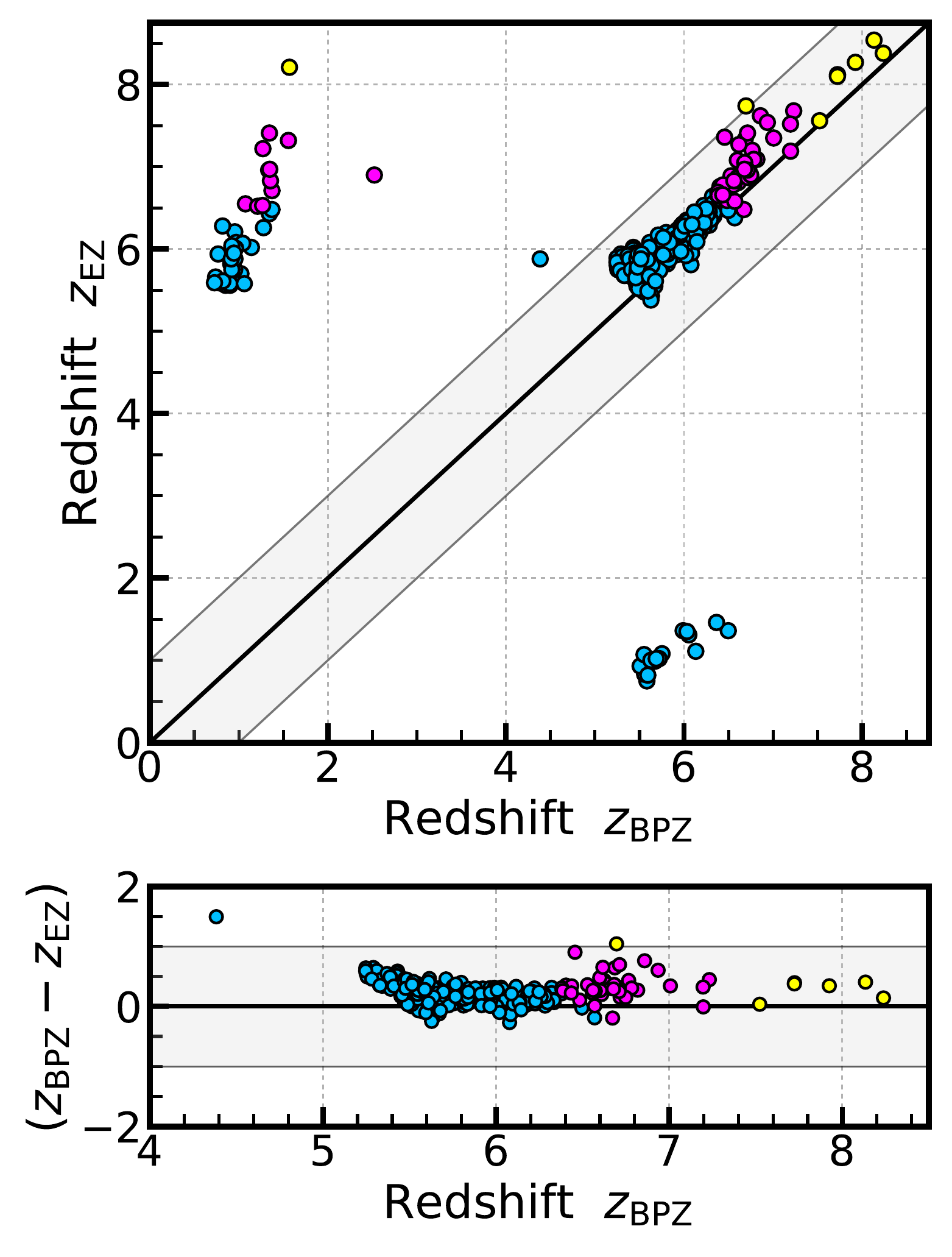}}
\caption{
Comparison of {EAZY} and {BPZ} photometric redshifts for our
high-$z$ sample. The grey regions show $\Delta z=\pm 1$. Objects with
large differences are due to one code preferring a $z\sim 1$ dusty or
high EW nebular emission line galaxy. Both {EAZY} and {BPZ} otherwise
agree within typical photometric-redshift uncertainties at these redshifts. 
}
\vspace{0.2 cm}      
\label{fig:Redshifts}
\end{figure} 
\begin{figure} \epsscale{1.1}       
\centerline{\includegraphics[scale=0.58]{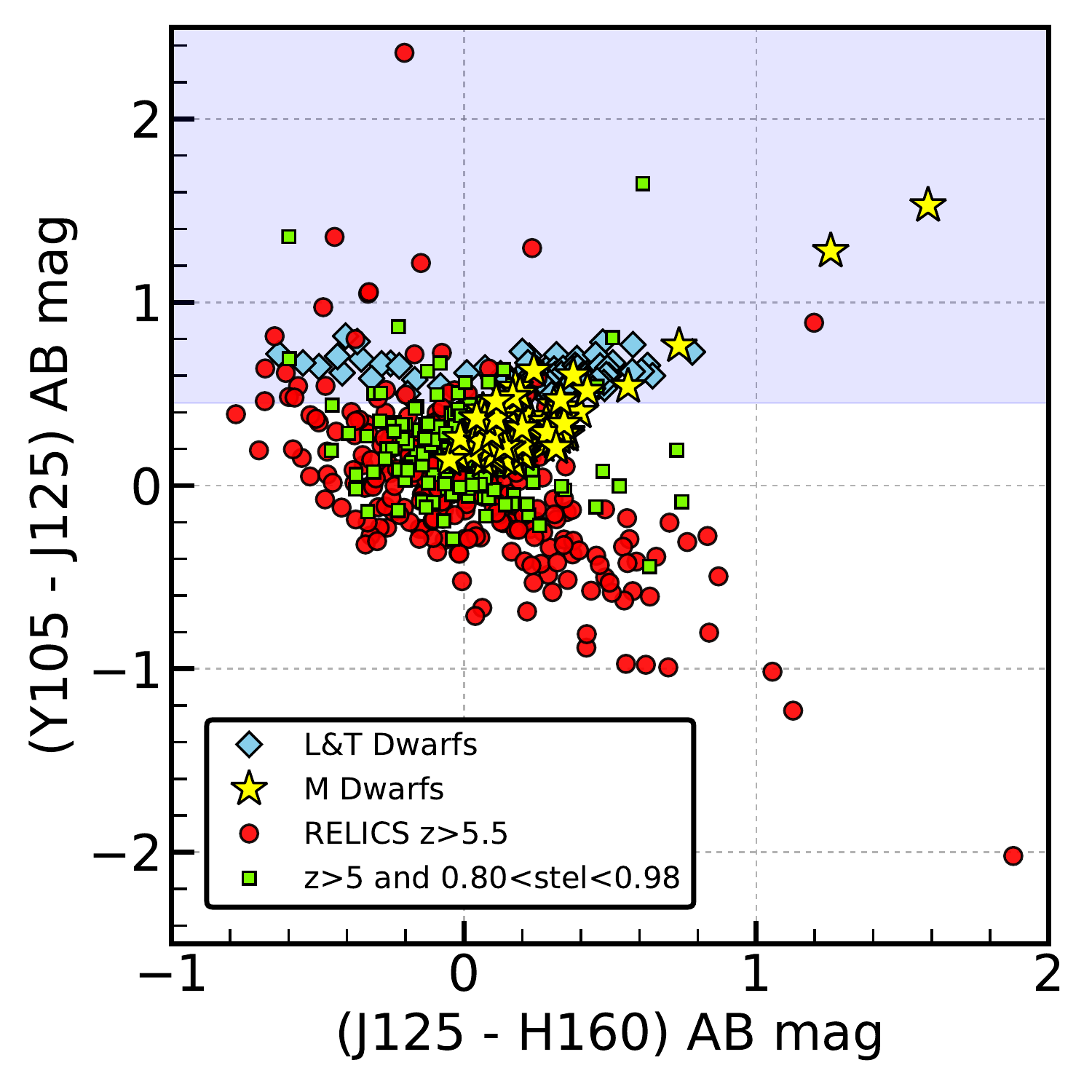}}
\caption{
The YJH colors of the RELICS high-$z$ galaxy candidates. The red
circles and green squares are objects with $z_{\rm phot, Max}>5.5$,
where the latter are those with high stellarity ($>0.8$). The blue
diamonds and yellow stars are colors of known L\&T and M dwarfs
respectively, taken from IRTF observed spectral library
\citep{Cushing05,Rayner09}.  Most of the high-redshift galaxy candidates with
high stellarity (green squares) have colors dissimilar from the dwarf
stars.  We remove all objects with $z_{\rm phot, Max}>5.5$,
$(Y-J)>0.45$, and stellarity $>$ 0.8, which correspond to the green
squares in the blue shaded region. 
}
\vspace{0.2 cm}      
\label{fig:StarColors}
\end{figure} 

We target all 41 clusters with two orbits of WFC3/IR comprising
observations in F105W, F125W, F140W, and F160W. Five clusters are
observed with an additional pointing, for a total of 46 IR fields. We
take advantage of existing archival ACS imaging, and for the 18
clusters without any F435W, F606W, and F814W we observe 3 orbits
total, with one orbit per filer. For the ACS imaging we also observe WFC3/IR fields in
parallel, which are not explored in this work. The observations are
split into two epochs separated by about a month to facilitate
variability search. Twenty additional orbits were allocated for variability
Target of Opportunity follow up.

The SExtractor \citep[version 2.8.6;][]{Bertin96} object selection and \hst\ photometry 
are described by Coe et al. (in prep), which we summarize here. First,
we use the AstroDrizzle package \citep{Gonzaga12} to combine all
sub-exposures from each filter. After aligning the filters to the same
pixel frame, we correct the absolute astrometry with the Wide-field Infrared
Survey Explorer (WISE) point source catalog \citep{Wright10}. Then, we
construct the final drizzled images by sampling the point-spread functions 
of both the ACS and WFC3/IR cameras in 30 milli-arcsecond (mas)/pixel and 60
mas/pixel scales. 

Finally, RELICS produces the full photometric catalogs of each cluster
field. In this work, we use the catalogs based on a detection image
comprised of the 60 mas weighted stack of all WFC3/IR imaging. 
The source extraction was performed with
SExtractor in dual-image mode, with fluxes measured within the
isophotal apertures. All fluxes are corrected for Galactic extinction
assuming the extinction law by \cite{Schlafly11}. 

\subsection{RELICS Spitzer Photometry}\label{sec:relicsPhot}
In addition to the new \hst\ imaging, RELICS also has tandem \spitzer\
IRAC programs (PI Brada{\v c}, PI Soifer) totaling 390 hours. The
IRAC 3.6~\micron\ and 4.5~\micron\ bands are especially helpful when calculating
photometric redshifts as their data helps to 
distinguish between $z>$5 galaxies and dusty $z\sim$2-3 galaxies. As these
data are still being processed, we do not yet employ the full
\spitzer\ photometry in the sample selection of this work. 
We will present the full RELICS \spitzer\ photometric catalogs in a future work. 

\subsection{Photometric Redshifts}\label{sec:Redshifts}
In this work, we utilize two different photometric-redshift fitting
codes to identify high-$z$ galaxy candidates: the Bayesian photometric
redshift code \citep[\foot{BPZ v1.99.3;}][]{
Benitez00,Benitez04,Coe06} and \foot{BPZ}and the Easy and Accurate
Z (photometric redshifts) from Yale
\foot{EAZY}\citep{Brammer08}. Both are similar in that they fit a
variety of empirically driven galaxy spectral energy distributions
(SEDs) to the data and find the template and redshift that best
matches the object. Photometric-redshift codes are fundamentally
similar to color-color selections in their identification of
high-redshift galaxies: they use input photometric bands to identify a
sharp increase in flux between two bands, and leverage with data at
other wavelengths to infer the presence of the Lyman break (due to the
line-of-sight neutral intergalactic medium absorbing photons at rest
wavelengths shorter than 1216~\AA), the Balmer break (which becomes
more pronounced in older stellar populations), or strong nebular
emission lines.  The main difference between photometric redshifts
and color-color selections is that the former is able to assign a
likelihood at each redshift by comparing the photometry to a library
of redshifted stellar population templates. We describe below the two
photometric-redshift codes and their implementation. 

\subsubsection{\foot{BPZ}redshifts}\label{sec:BPZ}
First, we derived photometric redshifts using \foot{BPZ.} 
\foot{BPZ}is based on $\chi^2$-fitting by comparing the
observed fluxes to PEGASE \citep{Fioc97} SED templates.
\foot{BPZ}employs a \cite{Madau95} intergalactic medium attenuation,
which accentuates the Lyman break. The default templates span a range
of rest-frame UVJ colors, and are combined to produce SEDs with
synthetic spectra similar to the high-quality spectra of most galaxies
\citep[e.g., with $\le$ 1~\% 
outliers, see][]{Coe13}, including red
dusty star-forming galaxies.  We use the default \foot{BPZ}templates
for fitting redshifts in this work.

Besides the choice of template SEDs, perhaps the biggest assumption in
a given photometric redshift code is the handling of the prior. In
\foot{BPZ,}the Bayesian prior $P(z,t|m_0)$ is a redshift and magnitude
($m_0$)-dependent prior applied to each template $t$, so as to
down-weight the likelihood of, for example, the unphysical presence of
bright elliptical galaxies at very high ($z>4$) redshifts. As we
discuss further in the following section, the exact prior for $z>6$ is
a poorly constrained function given that the population of
intrinsically faint galaxies of different types is unknown.
Nevertheless, we found that for \foot{BPZ}some simple prior must be
assumed to avoid an overpopulation of quiescent-like SEDs at high
redshift. With this prior, we calculate the posterior $P(z)$ for every
object in each RELICS field and define our accepted
\foot{BPZ}redshifts as the redshift corresponding to the mode of the
final probability function.

\subsubsection{\foot{EAZY}redshifts}\label{sec:EAZY}
Similar to \foot{BPZ,}the \foot{EAZY}photometric-redshift code creates the
redshift likelihood density function ($P(z)$) by computing the
$\chi^2$ between the observed fluxes and a linear combination of
redshifted empirical SED templates.  \foot{EAZY}includes 7 default
templates from PEGASE stellar population models \citep{Fioc97}, a red,
highly dust-obscured galaxy \citep{Maraston05}, and an extreme,
high-equivalent width (EW) nebular emission line galaxy 
\citep{Erb10} (see the Appendix Figure~\ref{fig:RestFrame} to compare the
rest-frame $UVJ$ and observed-frame $YJH$ colors of \foot{EAZY}and
\foot{BPZ}templates).  

While the $P(z)$ can be weighted by a $K$-band luminosity prior, we
choose to assume a flat prior for several reasons. First, we do not
wish to bias ourselves against high-$z$ galaxies that are unnaturally
bright due to high lensing magnification. Second, the priors at the
highest redshifts are poorly calibrated, as we are only just exploring
the completeness at these $z>6$ redshifts. More importantly, we 
input model fluxes of $z>6$ galaxies with typical data uncertainties and attempted to 
recover the redshift using the default \foot{EAZY}prior. We found 
that the default prior tends to systematically prefer the
high-EW low-$z$ solution over high-$z$ solutions. A better
estimation of the correct prior for each code may become more clear when
the photometric-redshifts are improved with the inclusion of the
Spitzer photometry.

For any photometric-redshift code, most uncertainty at $z>4$ redshifts
comes from the degenerate solutions with red, $z\sim1-2$ galaxies with
either high dust obscuration or evolved stellar populations. In the
Appendix, we show the WFC3 $YJH$ color tracks with increasing redshift
to show how high-$z$ galaxy colors are similar to $z=1-2$ red
galaxies.  
A slight preference to one of these degenerate solutions, due to the
different assumed templates, is the primary cause for the few cases where
\foot{BPZ}and \foot{EAZY}redshifts seem to differ by $\Delta z >1$. 

To be sure we were not omitting a population of high-$z$ galaxies that
systematically preferred the low-$z$ degenerate solution over the
high-$z$ (in either code), we visually inspected all galaxies by their 
morphologies, individual band images, stacked WFC3/IR and ACS
images, SEDs, and $P(z)$ distributions for all objects with appreciable
likelihood at high redshift, $P(z>4)> 40\%$. We concluded that there
were no convincing high-$z$ candidates (following the visual
inspection parameters described in S~\ref{sec:selection} below) that
did not have a maximal likelihood redshift or a median $P(z)$ redshift
of $z>5.5$ in at least one of the \foot{EAZY}or \foot{BPZ}fitting
results. 

To construct our sample of high-$z$ galaxy candidates we ultimately 
choose to adopt the average redshift between the \foot{BPZ}and \foot{BPZ}estimates 
unless  they differ by $\Delta z>1$ in which case we adopt the higher redshift solution. 
While this list will inevitable contain some low-$z$ contaminants, we aim to further validate 
the sample by a close inspection of the available \spitzer\ photometry or with follow-up
observations. 

\begin{figure*}[!t] 
\centerline{\includegraphics[scale=0.52]{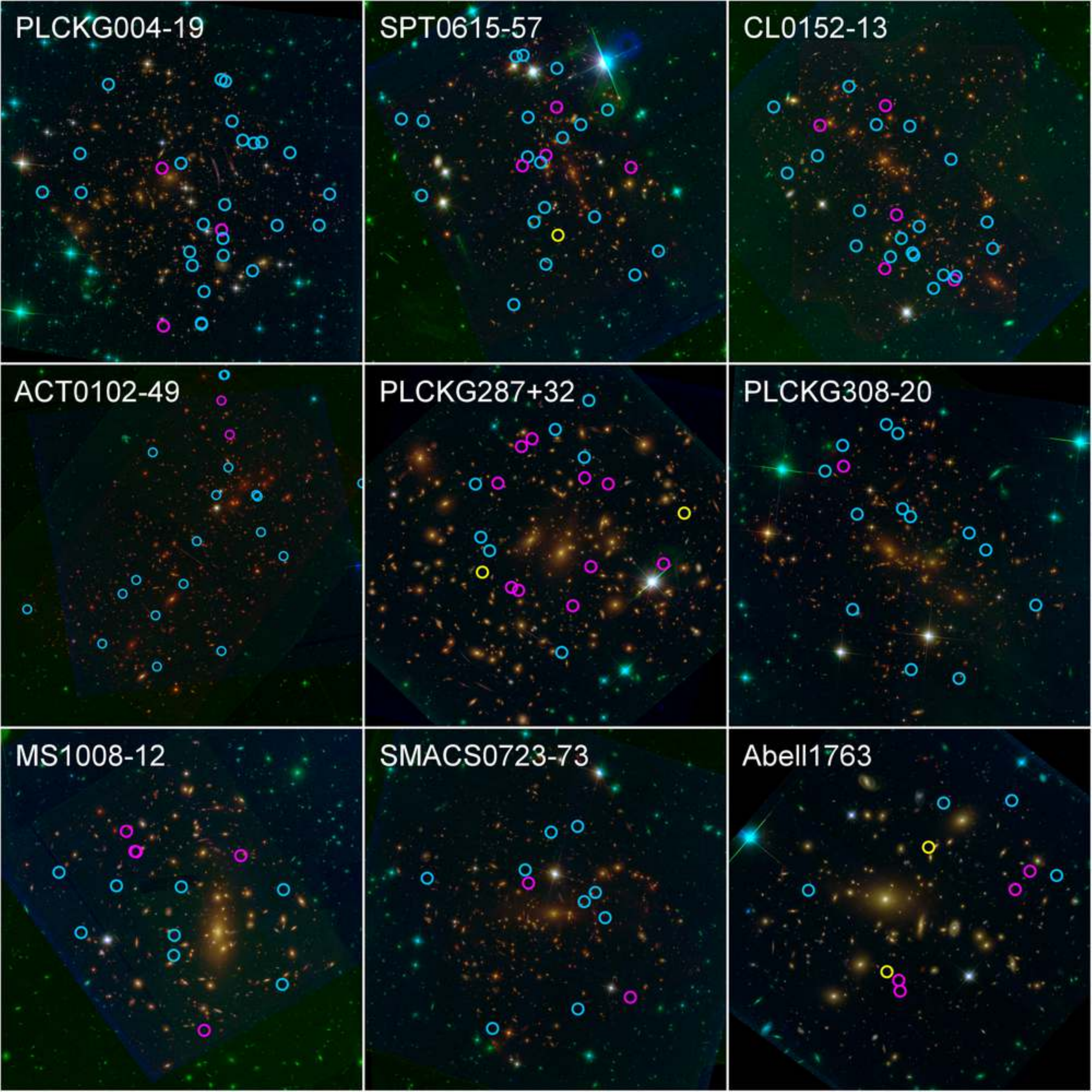}}
\caption{
Color images of the top nine clusters after rank ordering by the number of $z>5.5$ galaxies they
produce (excluding RXS0603+42 whose two WFC3/IR pointings are
separated by 6\arcmin). North is up and east is to the left. The images are scaled to
3\farcm 25 $\times$ 3\farcm 25, except for ACT0102-49 which is
4\farcm 25$\times$4\farcm 25. The cyan, magenta, and yellow circles mark the location of
the $z\sim$ 6, 7, and 8 candidate galaxies respectively. 
}
\vspace{0.2 cm}      
\label{fig:labeledclusters}
\end{figure*} 
\bgroup
\def\arraystretch{1.3}
\begin{table*}[!t]
\caption{High-$z$ Number Counts Per Cluster}
\label{tab:Clust}
\centering
\begin{tabular}{l r r c c c c c c }
\toprule
& R.A. & Dec. & Cluster & Planck Mass & & & & \\
Cluster & (J2000) & (J2000) & Redshift
& (10$^{14} \Msol$) & N$_{\rm total}$ & N$_{\rm z=6}$ & N$_{\rm z=7}$ & N$_{\rm z=8}$ \\
\hline
PLCKG004-19 & 19:17:4.50 & $-$33:31:28.5 & 0.520 & 10.36 & 28 & 25 & 3 & 0 \\
SPT0615-57 & 06:15:54.2 & $-$57:46:57.9 & 0.972 & 6.77 & 25 & 19 & 4 & 1 \\
CL0152-13 & 01:52:42.9 & $-$13:57:31.0 & 0.833 & . . . & 24 & 19 & 5 & 0 \\
ACT0102-49 & 01:02:53.1 & $-$49:14:52.8 & 0.870 & 10.75 & 21 & 19 & 2 & 0 \\
PLCKG287+32 & 11:50:50.8 & $-$28:04:52.2 & 0.39 & 14.69 & 19 & 7 & 10 & 2 \\
PLCKG308-20 & 15:18:49.9 & $-$81:30:33.6 & 0.480 & 10.32 & 14 & 13 & 1 & 0 \\
MS1008-12 & 10:10:33.6 & $-$12:39:43.0 & 0.306 & 4.94 & 13 & 8 & 5 & 0 \\
RXS0603+42 & 06:03:12.2 & +42:15:24.7 & 0.228 & 10.76 & 13 & 13 & 0 & 0 \\
SMACS0723-73 & 07:23:19.5 & $-$73:27:15.6 & 0.390 & 8.39 & 11 & 9 & 2 & 0 \\
Abell1763 & 13:35:18.9 & +40:59:57.2 & 0.228 & 8.13 & 10 & 4 & 4 & 2 \\
MACS0553-33 & 05:53:23.1 & $-$33:42:29.9 & 0.430 & 8.77 & 9 & 5 & 3 & 1 \\
MACS0257-23 & 02:57:10.2 & $-$23:26:11.8 & 0.505 & 6.22 & 9 & 8 & 1 & 0 \\
RXC0600-20 & 06:00:09.8 & $-$20:08:08.9 & 0.460 & 10.73 & 8 & 7 & 1 & 0 \\
MACS0025-12 & 00:25:30.3 & $-$12:22:48.1 & 0.586 & . . . & 7 & 6 & 1 & 0 \\
Abell2163 & 16:15:48.3 & $-$06:07:36.7 & 0.203 & 16.12 & 7 & 6 & 1 & 0 \\
Abell1758 & 13:32:39.0 & +50:33:41.8 & 0.280 & 8.22 & 6 & 6 & 0 & 0 \\
RXC0018+16 & 00:18:32.6 & +16:26:08.4 & 0.546 & 9.79 & 6 & 6 & 0 & 0 \\
Abell520 & 04:54:19.0 & +02:56:49.0 & 0.203 & 7.80 & 6 & 5 & 1 & 0 \\
MACS0308+26 & 03:08:55.7 & +26:45:36.8 & 0.356 & 10.76 & 6 & 6 & 0 & 0 \\
RXC0911+17 & 09:11:11.4 & +17:46:33.5 & 0.505 & 6.99 & 6 & 4 & 1 & 1 \\
Abells295 & 02:45:31.4 & $-$53:02:24.9 & 0.300 & 6.78 & 6 & 4 & 1 & 1 \\
Abell665 & 08:30:57.4 & +65:50:31.0 & 0.182 & 8.86 & 5 & 5 & 0 & 0 \\
Abell3192 & 03:58:53.1 & $-$29:55:44.8 & 0.425 & 7.20 & 5 & 2 & 3 & 0 \\
PLCKG209+10 & 07:22:23.0 & +07:24:30.0 & 0.677 & 10.73 & 5 & 5 & 0 & 0 \\
Abell2537 & 23:08:22.2 & $-$02:11:32.4 & 0.297 & 5.52 & 5 & 5 & 0 & 0 \\
SPT0254-58 & 02:54:16.0 & $-$58:57:11.0 & 0.438 & 9.69 & 4 & 3 & 1 & 0 \\
RXC0142+44 & 01:42:55.2 & +44:38:04.3 & 0.341 & 9.02 & 4 & 3 & 1 & 0 \\
Abell1300 & 11:31:54.1 & $-$19:55:23.4 & 0.307 & 8.97 & 4 & 2 & 2 & 0 \\
MACS0159-08 & 01:59:49.4 & $-$08:50:00.0 & 0.405 & 7.20 & 4 & 3 & 1 & 0 \\
MACS0035-20 & 00:35:27.0 & $-$20:15:40.3 & 0.352 & 7.01 & 4 & 3 & 1 & 0 \\
WHL0137-08 & 01:37:25.0 & $-$08:27:25.0 & 0.566 & 8.93 & 4 & 4 & 0 & 0 \\
Abell697 & 08:42:58.9 & +36:21:51.1 & 0.282 & 11.0 & 4 & 4 & 0 & 0 \\
PLCKG138-10 & 02:27:06.6 & +49:00:29.9 & 0.702 & 9.48 & 3 & 2 & 1 & 0 \\
PLCKG171-40 & 03:12:56.9 & +08:22:19.2 & 0.270 & 10.71 & 3 & 3 & 0 & 0 \\
RXC0032+18 & 00:32:11.0 & +18:07:49.0 & 0.396 & 7.61 & 3 & 2 & 1 & 0 \\
RXC0232-44 & 02:32:18.1 & $-$44:20:44.9 & 0.284 & 7.54 & 3 & 3 & 0 & 0 \\
RXC0949+17 & 09:49:50.9 & +17:07:15.3 & 0.383 & 8.24 & 3 & 3 & 0 & 0 \\
RXC1514-15 & 15:15:00.7 & $-$15:22:46.7 & 0.223 & 8.86 & 2 & 2 & 0 & 0 \\
RXC2211-03 & 22:11:45.9 & $-$03:49:44.7 & 0.397 & 10.5 & 2 & 2 & 0 & 0 \\
Abell2813 & 00:43:25.1 & $-$20:37:14.8 & 0.292 & 8.13 & 2 & 2 & 0 & 0 \\
MACS0417-11 & 04:17:33.7 & $-$11:54:22.6 & 0.443 & 12.25 & 0 & 0 & 0 & 0 \\
\hline
\end{tabular}
\end{table*}
\egroup

\subsection{High-$z$ Sample Selection} \label{sec:selection}   
The RELICS catalogs contain a combined total of over {76,000} sources.
From these sources we identify 2,425 objects with appreciable likelihood at $z>5.5$,
${P(z>5.5)>40\%}$. After initial visual inspections, we found that
galaxies only appeared to be bona-fide candidates (that is, S/N$>$3, small
sizes, and detections in individual infrared bands) if 
at least one of the photometric-redshift fitting codes had a median or
peak likelihood at $z>5.5$.  This lead us to adopting a single
redshift per object in order to produce a complete candidate
list. We took advantage of our use of two independent photometric
redshift codes by assigning the redshift of each object to be the average 
of the \foot{BPZ}and \foot{BPZ}estimates unless they differ by $\Delta z>1$ 
in which case we adopted the higher redshift solution. We note that the
\foot{BPZ}and \foot{EAZY}redshifts are in approximate agreement 
($|\Delta z|<1$) for 87\% of the sample.  We then selected objects with $z_{\rm phot}>5.5$
to reduce the initial candidate list to {1,337} objects.  We further refined the sample by selecting galaxies 
with an F160W detection greater than 3-$\sigma$ as adopted by \cite{Bradley14}, reducing to 841 objects. 

With any high-$z$ sample, we must be diligent to remove contamination
by foreground Galactic stars. L, T, and M dwarf stars and brown dwarfs
have similar broadband IR colors as $z=6-7$ galaxies \citep{Tilvi13}.
The SExtractor ``stellarity" parameter is one potentially effective way of doing so, 
with generally effective discriminatory power to $J~<~25$ or $J~<~26$ reported 
by \cite{Finkelstein15} and \cite{Bouwens15}, respectively, but with lesser 
reliability fainter than these levels. This is
exacerbated in lensing fields where high stellarity objects could be
stars or strongly lensed high-$z$ galaxies. To mitigate this problem,
we use both stellarity and a color selection to reduce our stellar
contamination.

Figure~\ref{fig:StarColors} shows the $YJH$ colors of the RELICS
high-$z$ sample compared to observed spectra of L, T, and M dwarf
stars \citep{Cushing05,Rayner09}. We outright remove objects with a
stellarity $\geq$98\% (111 objects) because these objects have very low FWHM
($<0.25 \arcsec$) and magnitudes systematically brighter than expected for
typical lensing magnifications (mean of 22.8 mag, well above the
distribution of magnitudes of lensed $z>$6 galaxies
from CLASH; see \cite{Bradley14}). Fig.~\ref{fig:StarColors}
shows that the colors of most objects at intermediate stellarity
(80-98\%) are dissimilar to stellar colors.  Until full lens models
are available for each field, we do not wish to rule out a
population of highly lensed, compact galaxies, especially given recent
estimates for the size of $2<z<8$ lensed galaxies in the HFF
\citep{Vanzella16a,Vanzella17,Bouwens17a}.  We therefore make the modest
selection to only omit high-$z$ objects that satisfy both high
stellarity (80-98\%) and $(Y-J)>0.45$ colors. 

Finally, we conduct an extensive visual inspection of all remaining
candidates, observing their detection in each band, ACS and WFC3
summed images, SExtractor segmentation maps, and the best-fit SEDs
and $P(z)$ from both \foot{BPZ}and \foot{EAZY.} The samples were
cleared of diffraction spikes, misidentified parts of larger galaxies,
stars, candidates too close to the infrared detector edge, transients
between epochs, and other image artifacts. For high-$z$ galaxies that
were obviously spatially distorted due to lensing, we remove the
duplicate segmentations (18 in total) and retain the brightest segment
to represent the object in the catalog. In total we find 255
galaxies at $z\sim 6$, 57 at $z\sim 7$, and 8 at $z\sim 8$.

\begin{figure}
\centerline{\includegraphics[scale=0.39]{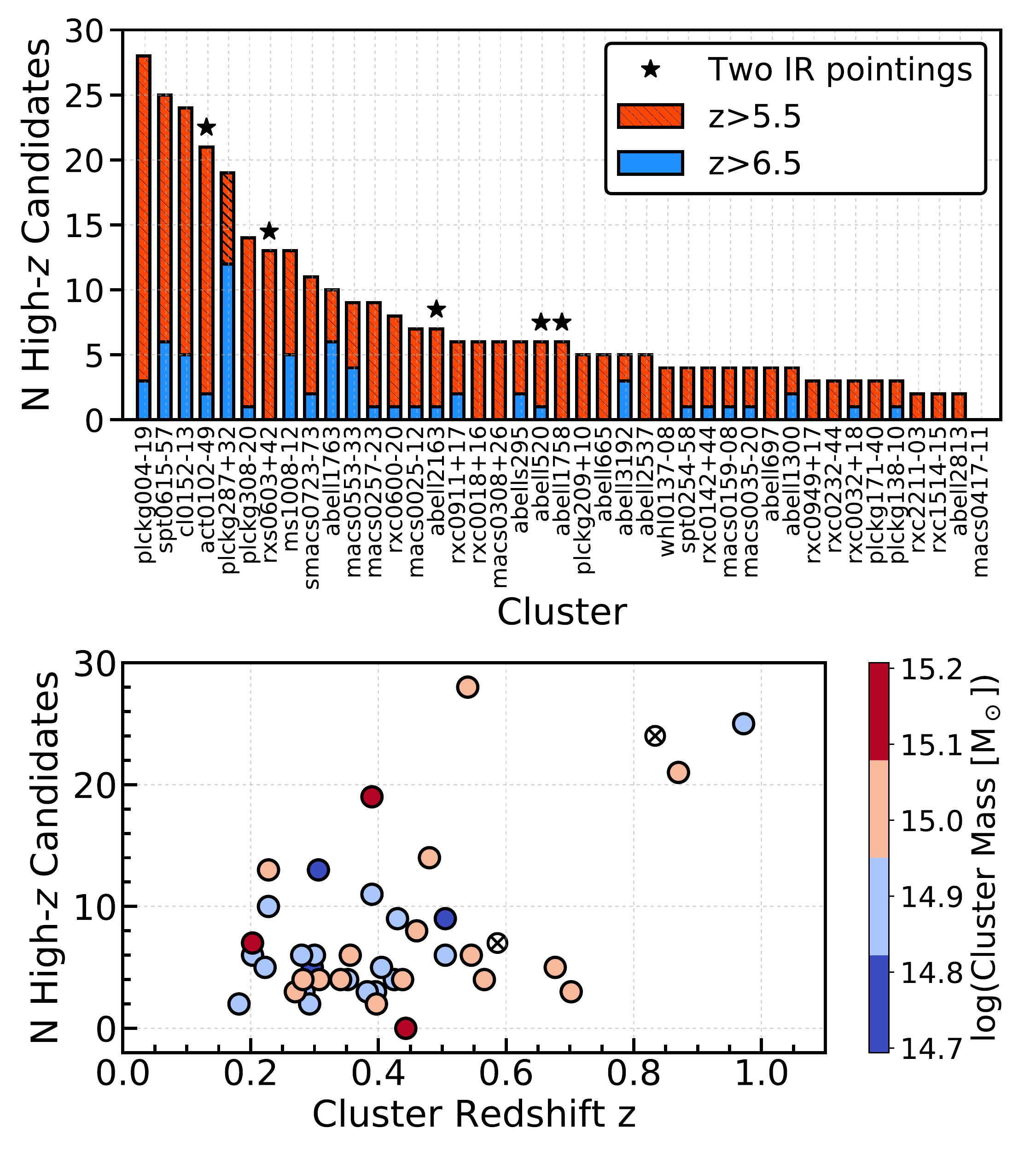}}
\caption{
The number of high-$z$ candidates for each cluster field observed by
RELICS. \emph{Top:} Histograms of the number of $z>5.5$ candidate
galaxies (solid blue and hatched red) and number of $z>6.5$ candidates
(solid blue only). The five clusters with two WFC3/IR pointings are
noted with the solid stars. Some clusters produce considerable numbers
of high-$z$ candidates compared to others. \emph{Bottom:} The number
of high-$z$ candidates per cluster as a function of the cluster
redshift. The blue-to-red colors portray the cluster mass
(M$_{500}$) from \cite{Planck16}, where the crossed circles show the
two clusters without mass estimates. There is a weak correlation
between the cluster redshift and the cluster's ability to produce many high-$z$
galaxy candidates, and little correlation with the cluster mass. 
}
\vspace{0.1 cm}      
\label{fig:ClustHist}
\end{figure} 
\begin{figure}[!t] 
\centerline{\includegraphics[scale=0.40]{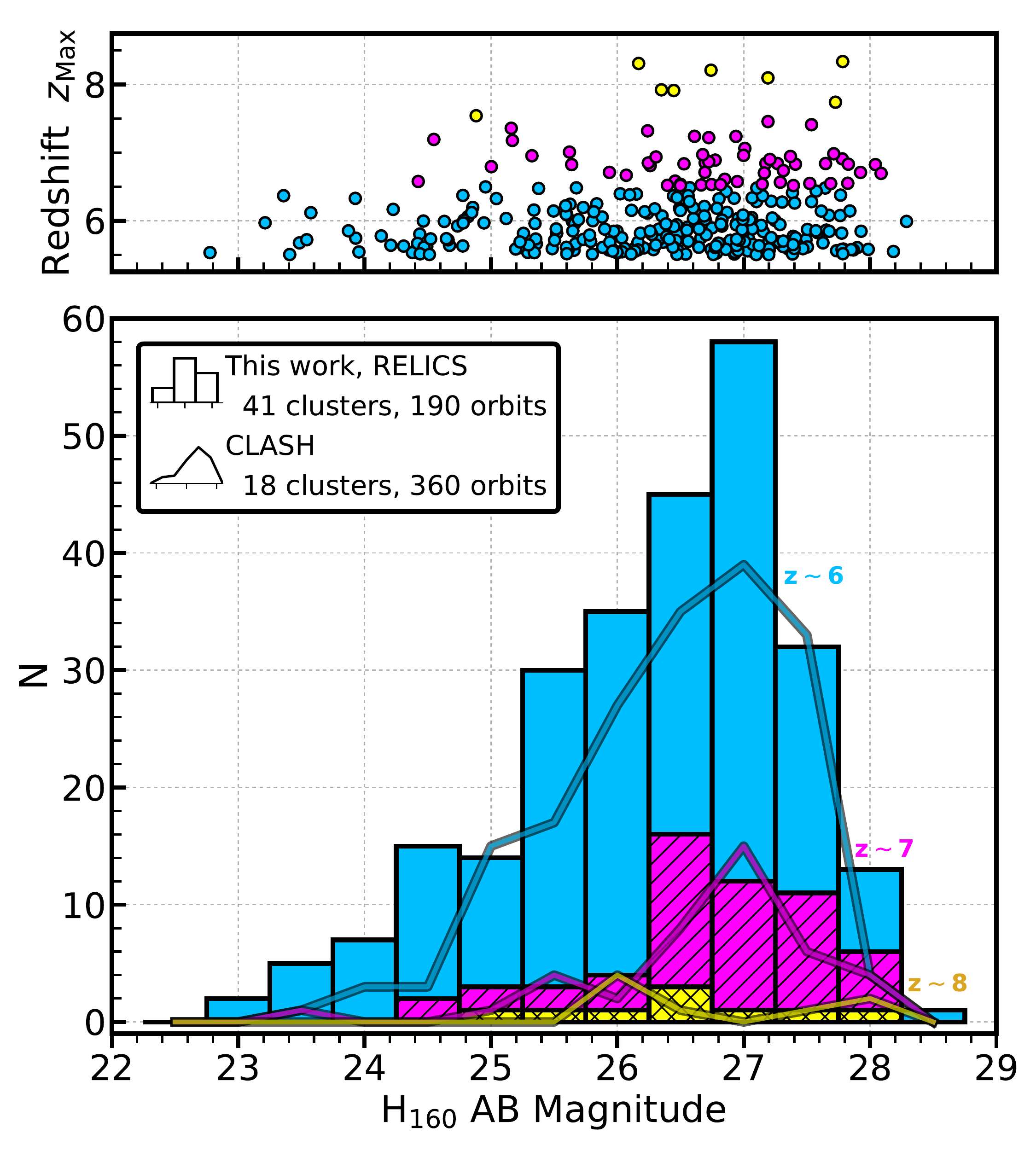}}
\caption{
The distribution of WFC3 F160W $H$-band observed (lensed) magnitude
for our $z\sim 6$, 7, and 8-10 RELICS samples in solid blue, left-hatched
magenta, and right-hatched yellow histograms. The curves show the
distribution of $z\sim$6, 7, and 8-10 galaxies (blue, magenta, and
yellow, respectively) from the first 18 clusters and 360 orbits of
CLASH \citep[][]{Bradley14}.  The above panel shows the photometric
redshifts of individual RELICS galaxies as a function of their
magnitude.  RELICS produces a similar magnitude distribution of high-$z$
galaxies as CLASH.
}
\vspace{0.2 cm}      
\label{fig:MagHist}
\end{figure} 

\section{Results}\label{sec:Results}
Figure~\ref{fig:labeledclusters} shows RGB color images of the top
nine high-$z$-producing fields (excluding RXS0603+42, whose two
WFC3/IR pointings are separated by 6\arcmin), and the overlaid
position of each high-$z$ candidate. These images show that our
candidate high-$z$ galaxies are not clustered around the edges of the
IR detector, thanks to our visual screening of every candidate. 
\cite{Cerny17} conducted an analysis of the first five RELICS
clusters, which span the range of masses and redshifts of the clusters
in the full program. They found that these five clusters had lensing
efficiencies of similar strength to the Frontier Fields. In future
works, we will publish the lens models and magnifications of all 41
clusters, which will allow us to explore the lensed counter images. 

The breakdown of the number of galaxies in each cluster and each
redshift bin are shown in Table~\ref{tab:Clust}.
Figure~\ref{fig:ClustHist} displays these number counts per cluster as
a histogram. Clearly, some clusters produce many more high-$z$
candidates than others, even after accounting for the five clusters
that have additional WFC3/IR pointings. For example, the top
$\approx$12\% of the high-$z$ producing clusters contain as many
candidates (almost a third of the entire $z>5.5$ sample) as the bottom
50\% of the clusters. Fig.~\ref{fig:ClustHist} also shows that there
appears to be little to no correlation between the ability of a
cluster to produce high-$z$ sources and its cluster mass, at least at
these high masses. However, a weak correlation exists with the number
of high-$z$ candidates and the cluster redshift. However, there may be
several reasons for why some clusters are better at producing high-$z$
sources than others, such as a dependence on the cluster mass
distribution and concentration, the latter of which is expected to
correlate with dark matter halo mass \citep{Bullock01,
Neto07,Umetsu14}. In addition, there may be sample variance in the
alignment of background galaxies. A comprehensive exploration of
cluster lensing strength will require a comparison between the lens
models from all RELICS clusters.

\subsection{Magnitude Distribution}\label{sec:MagDist}
Figure~\ref{fig:MagHist} shows distribution of F160W $H$-band
magnitude for the RELICS high-$z$ galaxies compared to that of the
CLASH survey.  RELICS produces the same, if not more, high-$z$
galaxies at a given redshift and magnitude than the first 360 orbits
of CLASH. The comparison to CLASH is important because although RELICS
viewed roughly twice as many clusters as CLASH, the infrared depth was
much shallower per cluster (typically a factor of 4 less IR exposure
time than CLASH) and the program used roughly a third as many orbits
to complete (compared to the total 524 \hst\ orbits by CLASH).
Fig.~\ref{fig:MagHist} also highlights the abundance of bright $m_{\rm
AB}<26$ candidates at a given redshift, which presents a promising
sample for follow-up spectroscopy. 

\begin{figure}[!t] 
\centerline{\includegraphics[scale=0.25]{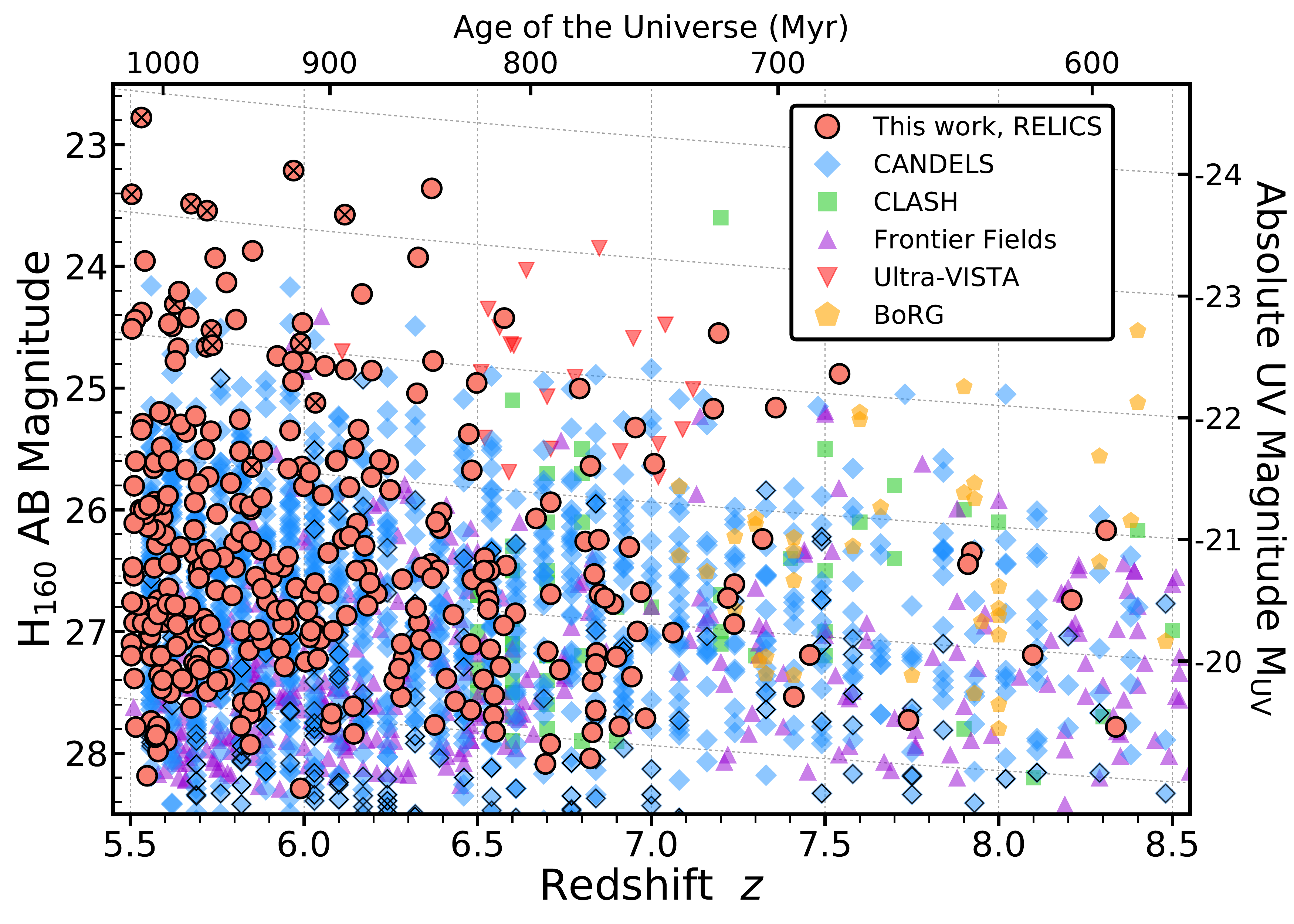}}
\caption{
The $H$-band magnitude as a function of redshift. The high-$z$
galaxy candidates from RELICS are shown as salmon-colored circles. Circles filled with an ``X"
mark candidates from the cluster field RXS0603+42, which is at a low Galactic latitude ($b\approx$10\degr) and therefore has a  potentially higher stellar contamination. The green squares are galaxies from CLASH  \citep{Bradley14}, the purple upwards-triangles from the Frontier Field \citep{Ishigaki17},
the red downwards-triangle from Ultra-VISTA \citep{Bowler17},
the blue diamonds from CANDELS \citep{Bouwens15} \citep[outlined diamonds are from the HUDF; see also][]{Finkelstein15}, and
the orange pentagons from BoRG/HIPPIES \citep{Bradley12,Schmidt14,Calvi16}.
Gray background lines follow the conversion from apparent magnitude to 
absolute UV magnitude. RELICS finds some of the brightest known galaxies at
a given redshift over $z\sim 8 $ to $z\sim 6$.
}
\vspace{0.2cm}
\label{fig:MagvZ}
\end{figure} 
\begin{figure*}
\centerline{\includegraphics[scale=0.36]{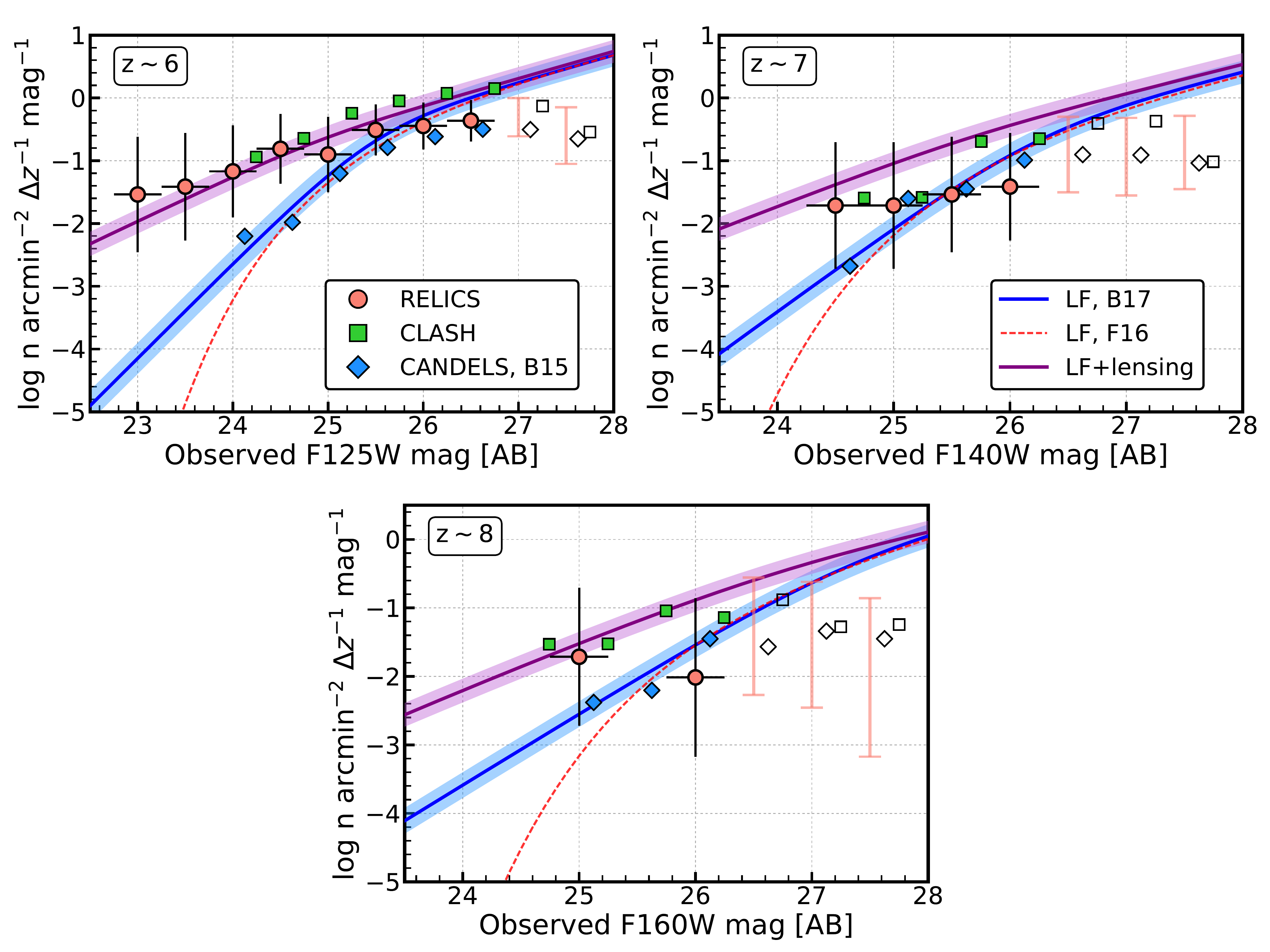}}
\caption{
The number density of galaxies per magnitude and redshift bin at
$z\sim$ 6, 7 and 8. The observed number densities in the 41-cluster RELICS
samples are shown as salmon-colored circles, with their 1-$\sigma$
Poissonian uncertainties. 
The green squares are the observed number densities seen in CLASH by
\cite{Bradley14} and the blue diamonds are those seen in CANDELS by
\cite{Bouwens15}.  The blue and red-dashed curves represent the
Double-Power-Law and Schechter fits to a suite of published luminosity 
function results \citep{Oesch12, Bradley12, McLure13, Bouwens15, 
Finkelstein15, Ono17, Stefanon17} from the literature (see Appendix A and
\cite{Finkelstein16}, respectively).   
The faint, capped error bars and open symbols show the number densities 
where faint-magnitude incompleteness begins to dominate. The purple
curves represent the expected number densities from CLASH after
simulating lensing effects on the \cite{Bouwens15} literature luminosity function.  
Compared to CLASH, RELICS yields similar number densities of $z\sim$6
galaxies, extending to brighter lensed magnitudes ($H\ m_{\rm AB}<26$).  
At $z\sim$7 and 8, RELICS yields somewhat lower number densities.
}
\vspace{0.2 cm}      
\label{fig:NumDens}
\end{figure*} 

Figure~\ref{fig:MagvZ} shows the $H$-band magnitude of the RELICS
high-$z$ candidates as a function of redshift compared to several
large surveys including CANDELS \citep{Bouwens15,Finkelstein15}, the
HFF \citep{Ishigaki17}, CLASH \citep{Bradley14}, Ultra-VISTA
\citep{Bowler17}, and BoRG/HIPPIES \citep{Bradley12,Schmidt14,Calvi16}.  
RELICS produces galaxies that are among the brightest at a given redshift over 
$z\sim 6$ to $z\sim 8$, comparable to these much wider and deeper programs.  
We highlight this comparison to emphasize the efficiency of targeting
strong lensing fields to produce high-$z$ candidates, which is
especially relevant as the costly overheads of \jwst\ make the
telescope more efficient at smaller area surveys. 

Finally, Figure~\ref{fig:NumDens} displays the number density of
galaxies in bins of magnitude over $z=$6-8. We assume an area of 4.5
arcmin$^2$ for each of the 46 WFC3/IR pointings for a total survey
area of 207 arcmin$^2$. The actual area will change after lens models
determine the magnification maps and the effective area covered for
each cluster, although typical areas from CLASH range between
${4.3 - 4.8}$ arcmin$^{2}$ per cluster field. Nevertheless, we observe
a clear excess in number density at the $z\sim 6$ and 7 bright
magnitudes compared to unlensed fields. We note that the drop off at
fainter magnitudes ($m_{\rm AB}>26.5$) is due to survey
incompleteness. 

To provide a baseline comparison for our results, we make use of a LF with 
a double power-law fit to a comprehensive set of $z\sim 6$, 7, and 8 results 
from the literature (see Appendix).   A double power-law has been found to 
work well in representing the extreme bright-end shape of the UV LF 
\citep{Bowler14, Bowler15, Ono17, Stefanon17}.   By comparing the bright end of our results 
from RELICS against that of a double-law LF, we aim to show that the number density
boost from lensing we observe exceeds even the expectation from the double-law LF, 
which already exhibits a much larger number of sources than a LF with a Schechter 
\citep{Schechter76} form.  As a second baseline comparison for our results, we also show 
the expectations from  the literature-averaged Schechter function results from \cite{Finkelstein16} which features 
fewer sources at the extreme bright-end due the bright-end shape of the Schechter function, 
but overall is very similar at luminosities less than 2$L^*$. 

Next, we apply typical CLASH cluster-lensing magnifications to model how the number
densities of the true luminosity function appear under the effects of
cluster lensing. We only show the effects of lensing on the double-law LF, but we note that
forms of the LF produce a similar lensed shape. Comparing the lensed luminosity function 
to the binned number densities of RELICS and CLASH, we find they agree with the prediction 
well at $z\sim 6$, and that both surveys tend to underproduce from the expected number of sources at
$z\sim 7$ and 8. It is hard to speculate if this is due to an actual bright-end decline 
in number density at higher redshifts, as seen by  \cite{Bowler17}, until we conduct 
full completeness simulations and lens modeling of all 41 RELICS clusters and 
consider the effects of cosmic variance. 

\subsection{Exceptionally Bright Sources}\label{sec:Bright}
The RGB image stamps of the brightest 40 galaxy candidates from RELICS
for $z\sim 6$ and 7, and all candidates for $z\sim$8 are shown in Figures~\ref{fig:z6stamps}, \ref{fig:z7stamps}, \ref{fig:z8stamps} respectively. In particular, we note the third brightest $z\sim6$ candidate, MACS0308-904, which has been clearly arced by the effects of lensing and is likely highly magnified. Not all highly magnified galaxies will also be arced, which makes it very difficult with the current data to distinguish between stellar contaminants and high-$z$ galaxies among the brightest candidates. We have already attempted to remove stellar contaminants by a combination of YJH colors and stellarity (Fig.~\ref{fig:StarColors} and \S~\ref{sec:selection}). In addition, we checked the brightest candidates in our samples with the Galactic latitude of their cluster field. We find that $\approx$6 of our brightest $z\sim6$ candidates come from a cluster field with relatively low Galactic latitude (RXS0603+42, at $b$= 9.7\degr). We specially note these objects in Fig.~\ref{fig:MagvZ} and their images and SEDs can be inspected in Fig.s~\ref{fig:z6stamps} and \ref{fig:z6SEDs}. Besides this one cluster, we see no correlation between the number or brightness of high-$z$ candidates and the Galactic latitude of the cluster fields. We anticipate a deeper exploration of contaminants in the future using lensing magnifications, \spitzer\ photometry, and/or spectroscopic redshifts. 

Following the same plotting grid of the image stamp figures, the SEDs of
the brightest candidates and their photometric-redshift
template fits for \foot{BPZ}and \foot{EAZY}are shown in 
Figures~\ref{fig:z6SEDs}, \ref{fig:z7SEDs}, \ref{fig:z8SEDs}
respectively. Several objects appear to have red SEDs, but we caution that this is because the rest-frame optical is unconstrained prior to the inclusion of \spitzer, which makes estimates of physical parameters like stellar mass unreliable. In Tables \ref{tab:z6}, \ref{tab:z7}, and \ref{tab:z8} we make available the $z$= 6, 7, and 8 candidate galaxies from the RELICS survey. We include both the photometric estimates of \foot{EAZY}and \foot{BPZ.} The full tables will be made available online.

\section{Conclusions}\label{sec:Conclusions}
We present the candidate high-$z$ galaxies first estimated from 
RELICS, an \hst\ Treasury Program observing 41 galaxy clusters. We use two
independent photometric-redshift fitting codes to determine the
redshifts of each galaxy. We also compare the colors of the candidates
to those of known dwarf stars, and apply a color selection to remove
the most likely contaminants. Furthermore, we conduct an extensive visual
inspection of all potential high-$z$ candidates, cleaning the sample
of diffraction spikes, misidentified parts of larger galaxies,
stars, spurious noise close to the infrared detector edge, transients
between epochs, and other image artifacts. 

The final sample of candidate high-$z$ galaxies is of comparable size
to the CLASH program, despite being significantly more shallow. In
particular, we identify several candidates that are among the
brightest galaxies at $z\sim$6 to 8, as compared to much deeper and wider area
surveys. This presents a promising sample for follow-up
spectroscopy to study the nebular ionization conditions,
Lyman-$\alpha$ emission, and other galaxy properties into the epoch 
of reionization. 

Finally, we compare the number of galaxies found to the predictions
from lensed luminosity functions. We find that our $z\sim 6$ sample agrees
with expectations from the literature luminosity functions, while
$z\sim$7 and 8 begin to under-produce the number of candidates compared 
to expectations from current luminosity functions. The paucity of galaxies from
the higher redshift samples could imply the start of an accelerated bright-end evolution
in the UV LF, although we await the completion of the \spitzer\ data and completeness 
simulations for further confirmation. 
 

\section*{Acknowledgements} 
We thank Gabriel Brammer for insightful discussions related to this
work.  This paper is based on observations made with the NASA/ESA
Hubble Space Telescope.  The Space Telescope Science Institute (STScI)
is operated by the Association of Universities for Research in
Astronomy, Inc. (AURA) under NASA contract NAS 5- 26555. ACS was
developed under NASA contract NAS 5-32864. The Spitzer Space Telescope
is operated by the Jet Propulsion Laboratory, California Institute of
Technology under a contract with NASA.  These observations are
associated with program GO-14096.  Archival data are associated with
programs GO-9270, GO-12166, GO-12477, GO-12253.  Some of the data
presented in this paper were obtained from the Mikulski Archive for
Space Telescopes (MAST).  This work was performed under the auspices
of the U.S. Department of Energy by Lawrence Livermore National
Laboratory under Contract DE-AC52-07NA27344. F.A.-S. acknowledges
support from Chandra grant G03-14131X.

\bibliographystyle{apj} \bibliography{salmon17}

\clearpage

\appendix
\section{The Bright end of the UV Luminosity Function}

The regions on the sky immediately surrounding RELICS clusters will
benefit from substantial amounts of lensing magnification and
therefore should benefit from a significant enhancement in the surface
density of especially bright, highly magnified sources.  While one
would expect more bright sources in lensing fields relative to Schechter
representations of the $UV$ LF at $z\sim6$-8, we must take notice that
the consensus of the un-lensed bright end $UV$ LF has changed in recent years. 
Thanks to recent wide-field surveys, there is now increasingly compelling evidence 
that the bright end of the $z\sim6$-8 $UV$ LF
exhibits more of a double power-law form \citep{Bowler14, Bowler15, Ono17}:
\begin{equation}
\phi^* \left[(\frac{L}{L^*})^{-\alpha} + (\frac{L}{L^*})^{-\gamma} \right ]^{-1} \frac{dL}{L^*}
\end{equation}
Compared to a relatively abrupt bright-end cut-off
inherent to Schechter function fits, the double power-law form of the LF
accurately recovers the larger number of observed bright sources and plausibly yields
a similar number of bright sources to what we would expect based on
lensing magnification from our RELICS sample.

Therefore, to demonstrate more clearly the gains one achieves in
identifying bright sources behind lensing clusters (vs. blank field
searches), we make use of double power-law representations of the LF
results in the literature.  In deriving these double power-law LF
results, we make use of LF constraints from the Hubble Ultra Deep
Field, the Hubble Ultra Deep Field parallel fields, Hubble Frontier
Field parallel fields, CANDELS, UltraVISTA, UDS, and the
Hyper-Suprime-Cam fields.  We executed these fits using a Markov-chain
Monte-Carlo procedure.

At $z\sim6$, we make use of the published stepwise LF constraints from
\cite{Bouwens07, Bouwens15, Finkelstein15, Bowler15, Ono17}; at $z\sim7$, we
make use of LF constraints from \cite{McLure13, Bouwens15, Finkelstein15, 
Bowler17, Ono17}; at $z\sim8$, we make use of LF constraints from 
\cite{Oesch12, Bradley12, McLure13, Bouwens15, Finkelstein15, Stefanon17}.

For LF constraints derived from Lyman-break selections, where the mean
redshift is less than that for photometric redshift selections, we have
adjusted the published volume densities to the expected
differences relative to the LF results at $z=6$, $z=7$, and $z=8$.
Specifically, we adjusted the \cite{Bouwens15} $z=5.9$, $z=6.8$,
and $z=7.9$ downwards by 0.02 dex, 0.04 dex, and 0.02 dex,
respectively, and \cite{Bouwens07} $z=5.9$ results downwards by
0.02 dex.

A few points in the aforementioned LF determinations from the
literature appear to be subject to sizable systematics, and are 
therefore excluded from our LF fits. Among these is the 
faintest stepwise point in the $z\sim7$ LF from \cite{Bowler17} 
which is discrepant by several sigma from the other LF determinations. 
It is likely that this point is impacted by uncertainties in the
estimated completeness at the faint end of UltraVISTA, and not due to 
the start of a shallower faint-end slope. 
Also excluded are the intermediate-luminosity (i.e., $-20$ to $-19$)
$z\sim6$ and $z\sim7$ stepwise points from \cite{Finkelstein15},
which receive a dominant contribution from search results over
CANDELS. These points were excluded because the faint-end corrections for completeness 
over the CANDELS fields may have been overestimated, which accounts for the differences between 
the results by \cite{Finkelstein15} and the results from other fields , such as \cite{Bouwens15} and \cite{McLure13}.

\begin{figure*}
{\centerline{\includegraphics[width=7.1in,height=3in,trim={20pt 350pt 30pt 220pt},clip=true]{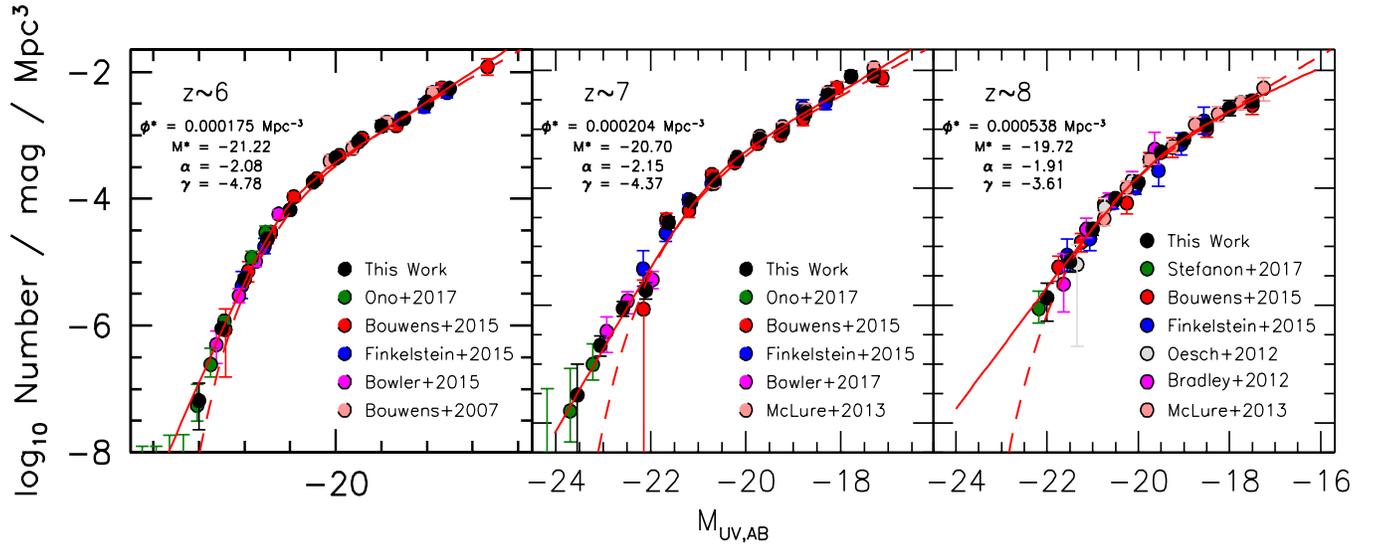}}}
\caption{
Best-fit double power-law determinations of the $z\sim6$, $z\sim7$,
and $z\sim8$ LF results using a comprehensive set of published LF
results from the literature (solid circles).  Best-fit LF parameters
are included in each panel, while the solid red lines show the
double-power law fits.  Included in the $z\sim6$ fits are the
determinations from \cite{Bouwens07, Bouwens15, Finkelstein15, Bowler15, Ono17}; 
included in the $z\sim7$ fits are determinations from 
\cite{McLure13, Bouwens15, Finkelstein15, Bowler17, Ono17}; included in the
$z\sim8$ fits are determinations from \citep{Oesch12, Bradley12, McLure13, 
Bouwens15, Finkelstein15, Stefanon17}.  Slight adjustments
(10-20\%) have been made to the volume density of individual points to
account for slight differences in the mean redshifts of the derived LF
results.  The faintest constraint on the $z\sim7$ LF from \cite{Bowler15} 
and several intermediate luminosity LF constraints from
\cite{Finkelstein15} due to concerns they likely overestimate
the completeness of the UltraVISTA and CANDELS selections.  The
\cite{Finkelstein16} literature-averaged Schechter fits are shown with
the dashed lines.
}
\vspace{0.2 cm}      
\label{fig:Bouwens}
\end{figure*}

The best-fit double-law LF results we obtain are presented in Figure~\ref{fig:Bouwens}
with the red solid lines.  The best-fit double power-law parameters
are $\phi^*=0.000175$ Mpc$^{-3}$, $M^* = -21.22$, $\alpha=-2.08$,
$\gamma=-4.78$ at $z\sim6$, $\phi^*=0.000204$ Mpc$^{-3}$, $M^* =
-20.70$, $\alpha=-2.15$, $\gamma=-4.37$ at $z\sim7$, $\phi^*=0.000538$
Mpc$^{-3}$, $M^* = -19.71$, $\alpha=-1.91$, $\gamma=-3.61$ at
$z\sim8$.  As the present results also represent fits to LF results in
the literature, our fit results are in reasonable agreement with
previous literature-averaged fits by \cite{Finkelstein16} using a
Schechter functional form at all luminosities except the extreme
bright end.

\begin{figure*}[!h]
\centerline{\includegraphics[scale=0.49]{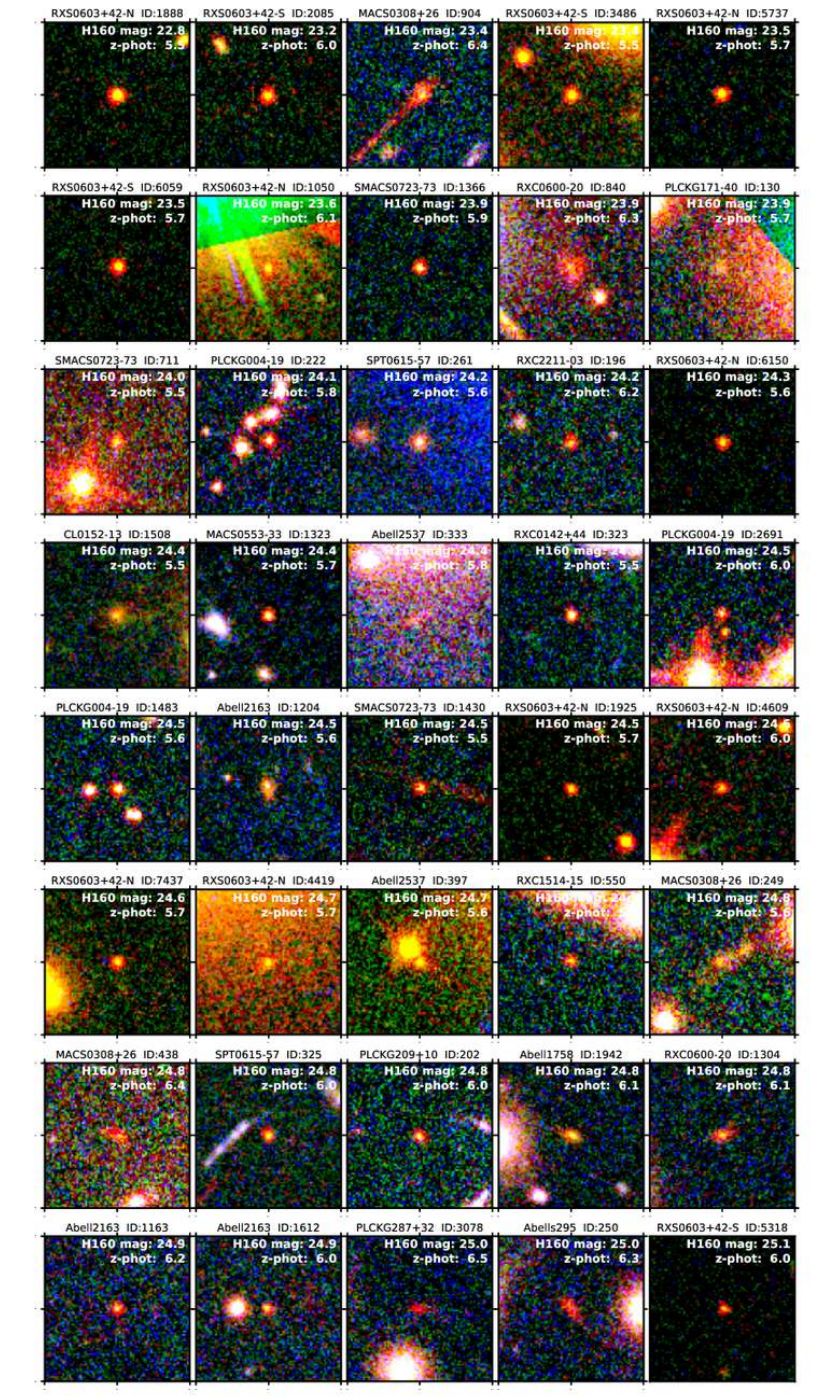}}
\caption{
The brightest 40 galaxy candidates from the $z\sim6$ RELICS sample.
Each RGB color image stamp is 5\arcsec x5\arcsec\ with the red channel
as the sum of all IR bands, the G channel as the ACS F814W band, and
the B channel the sum of ACS F435W and F606W.  The F160W $H$-band AB
magnitude is shown within each stamp, along with the adopted redshift (see \S~\ref{sec:selection}). The
cluster name and catalog ID are shown at the top of each stamp. 
}
\vspace{0.2 cm}      
\label{fig:z6stamps}
\end{figure*} 
\begin{figure*}[!h]
\centerline{\includegraphics[scale=0.49]{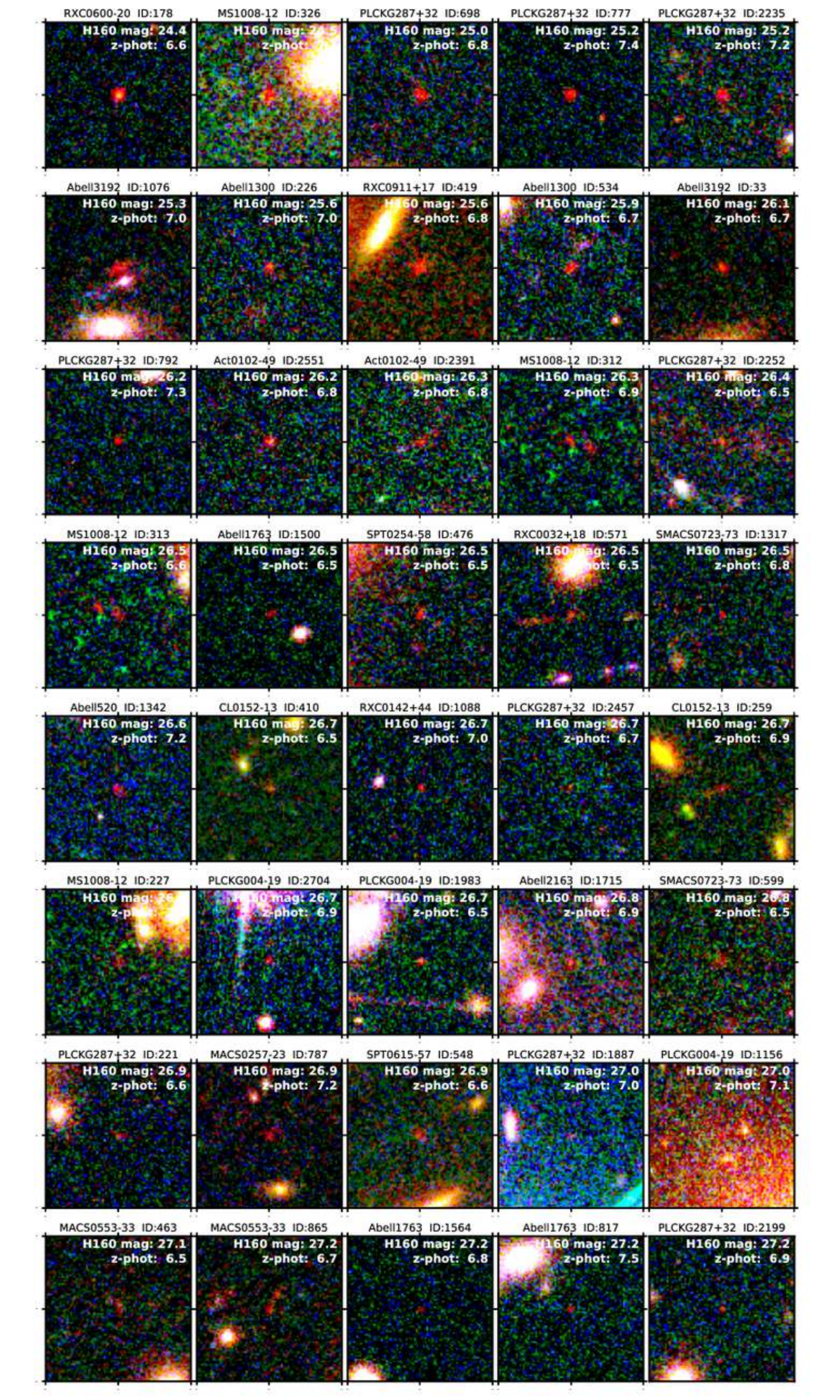}}
\caption{
The same as Fig.~\ref{fig:z6stamps} but for the brightest 40 objects
from the $z\sim7$ sample. 
}
\vspace{0.2 cm}      
\label{fig:z7stamps}
\end{figure*} 
\begin{figure*}[!h]
\centerline{\includegraphics[scale=0.4]{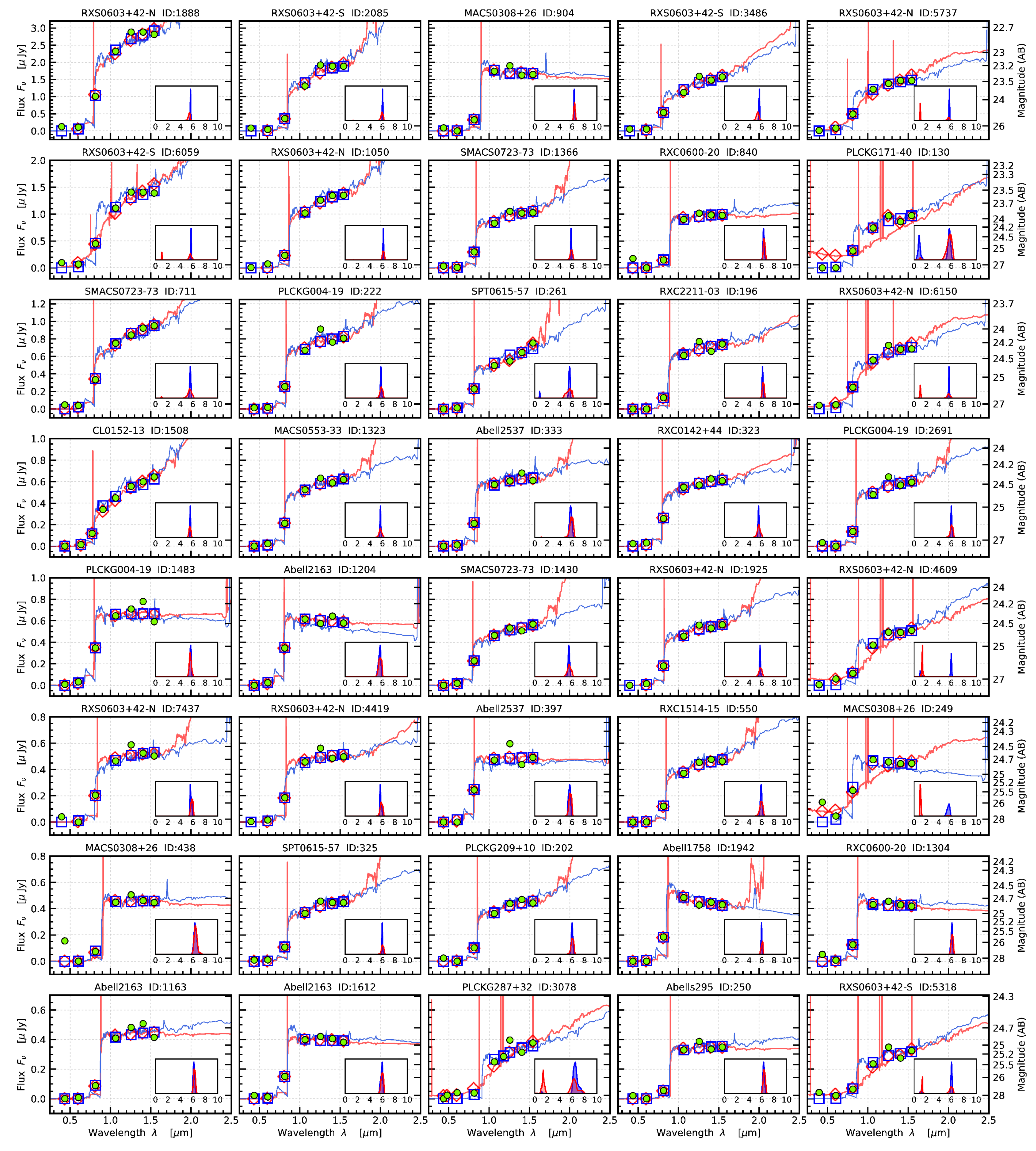}}
\caption{
The best-fit SEDs from {EAZY} (red) and {BPZ} (blue) for the 40 brightest
galaxy candidates from the $z\sim6$ RELICS sample. The
diamonds and squares show the expected fluxes from the model SED for
{EAZY} and {BPZ} respectively, and the green circles are the RELICS data.
The typical photometric uncertainties are about the size of the data points. 
The top title shows the cluster and catalog ID number of each high-$z$
candidate. The grid of this plot matches to the galaxies shown in
Fig.~\ref{fig:z6stamps}.
}
\vspace{0.2 cm}      
\label{fig:z6SEDs}
\end{figure*} 
\begin{figure*}[!h]
\centerline{\includegraphics[scale=0.4]{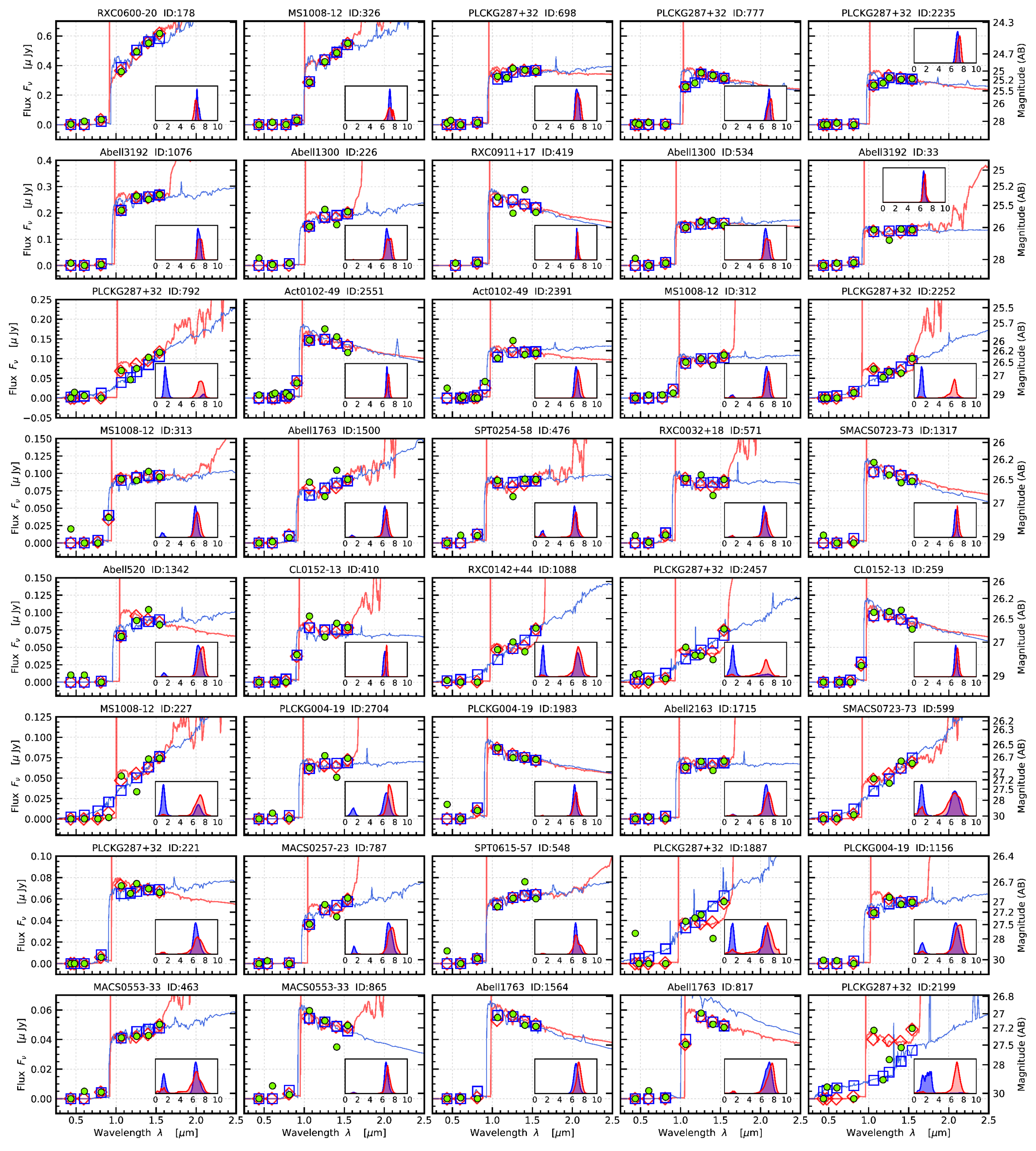}}
\caption{
The same as Fig.~\ref{fig:z6SEDs} but showing the 40 brightest
galaxies from the $z\sim7$ RELICS sample. The grid of this plot
matches to the galaxies shown in Fig.~\ref{fig:z7stamps}.
}
\vspace{0.2 cm}      
\label{fig:z7SEDs}
\end{figure*} 
\begin{figure*}[!h]
\centerline{\includegraphics[scale=0.60]{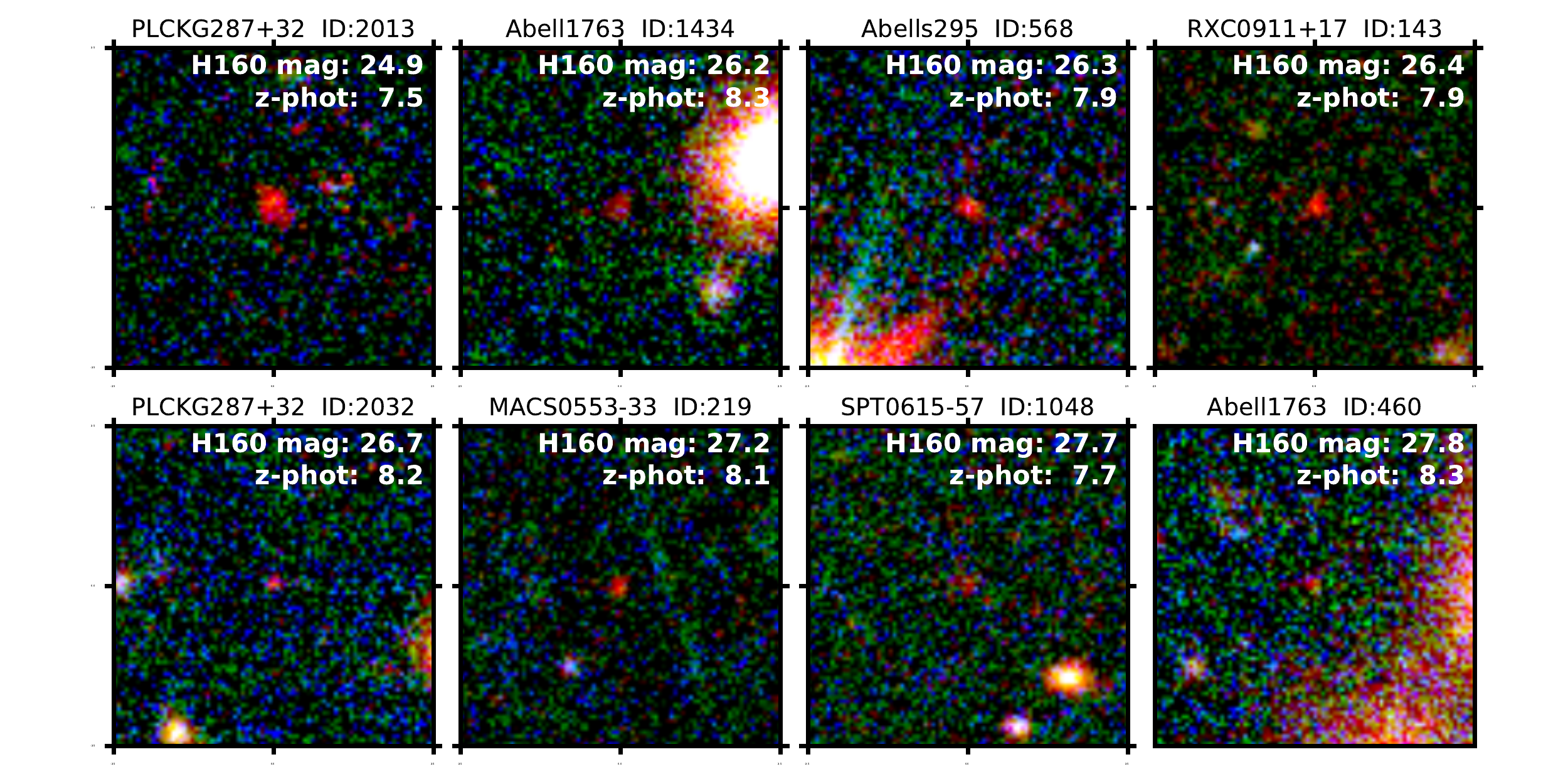}}
\caption{
All galaxy candidates from the $z\sim8$ RELICS sample.
Each RGB color image stamp is 5\arcsec x5\arcsec\ with the red channel
as the sum of all IR bands, the G channel as the ACS F814W band, and
the B channel the sum of ACS F435W and F606W.  The F160W $H$-band AB
magnitude is shown within each stamp, along with the the adopted 
redshift (see \S~\ref{sec:selection}). The
cluster name and catalog ID are shown at the top of each stamp. 
}
\vspace{0.2 cm}      
\label{fig:z8stamps}
\end{figure*} 
\begin{figure*}[!h]
\centerline{\includegraphics[scale=0.42]{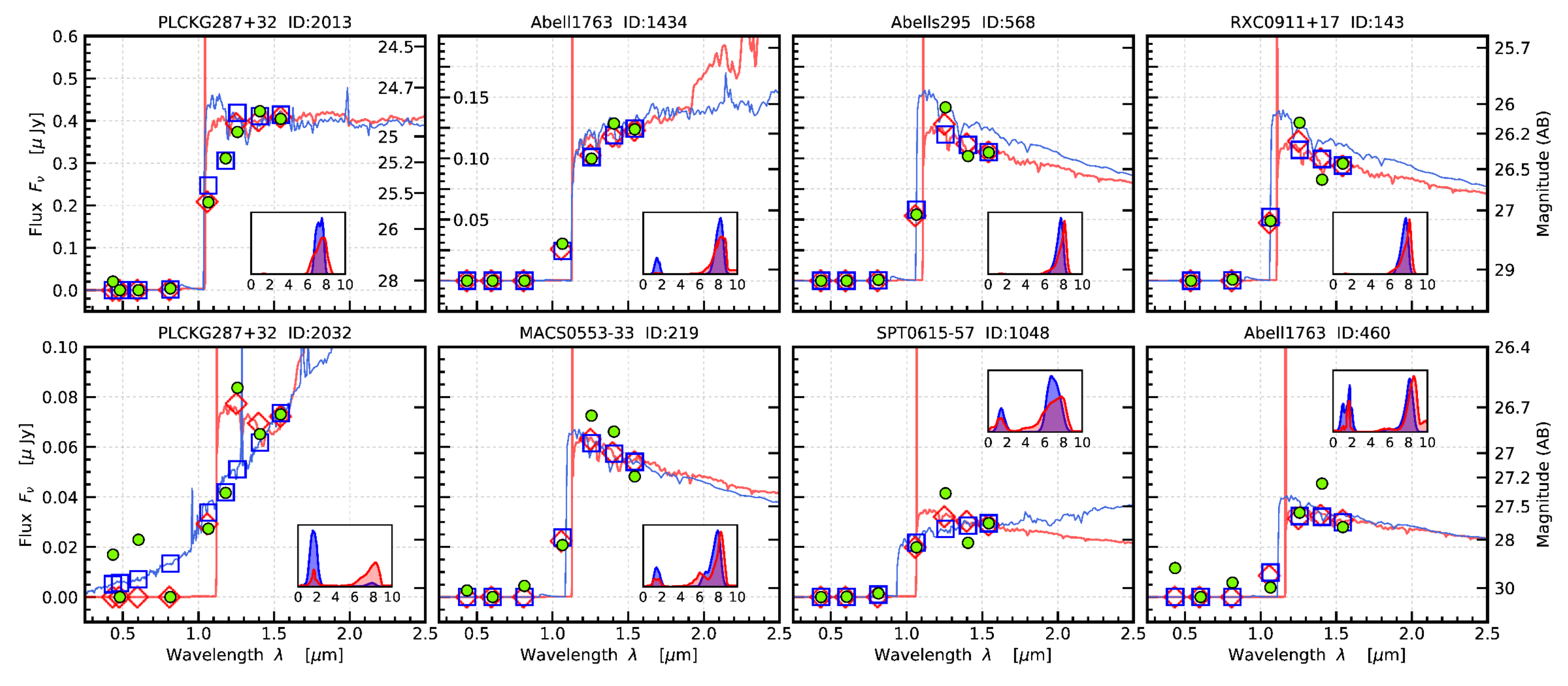}}
\caption{
The best-fit SEDs from {EAZY} (red) and {BPZ} (blue) for all 
galaxy candidates from the $z\sim8$ RELICS sample. The
diamonds and squares show the expected fluxes from the model SED for
{EAZY} and {BPZ} respectively, and the green circles are the RELICS data.
The typical photometric uncertainties are about the size of the data points. 
The top title shows the cluster and catalog ID number of each high-$z$
candidate. The grid of this plot matches to the galaxies shown in
Fig.~\ref{fig:z8stamps}.
}
\vspace{0.2 cm}      
\label{fig:z8SEDs}
\end{figure*} 

\clearpage

\begin{figure*}
\centerline{\includegraphics[scale=0.80]{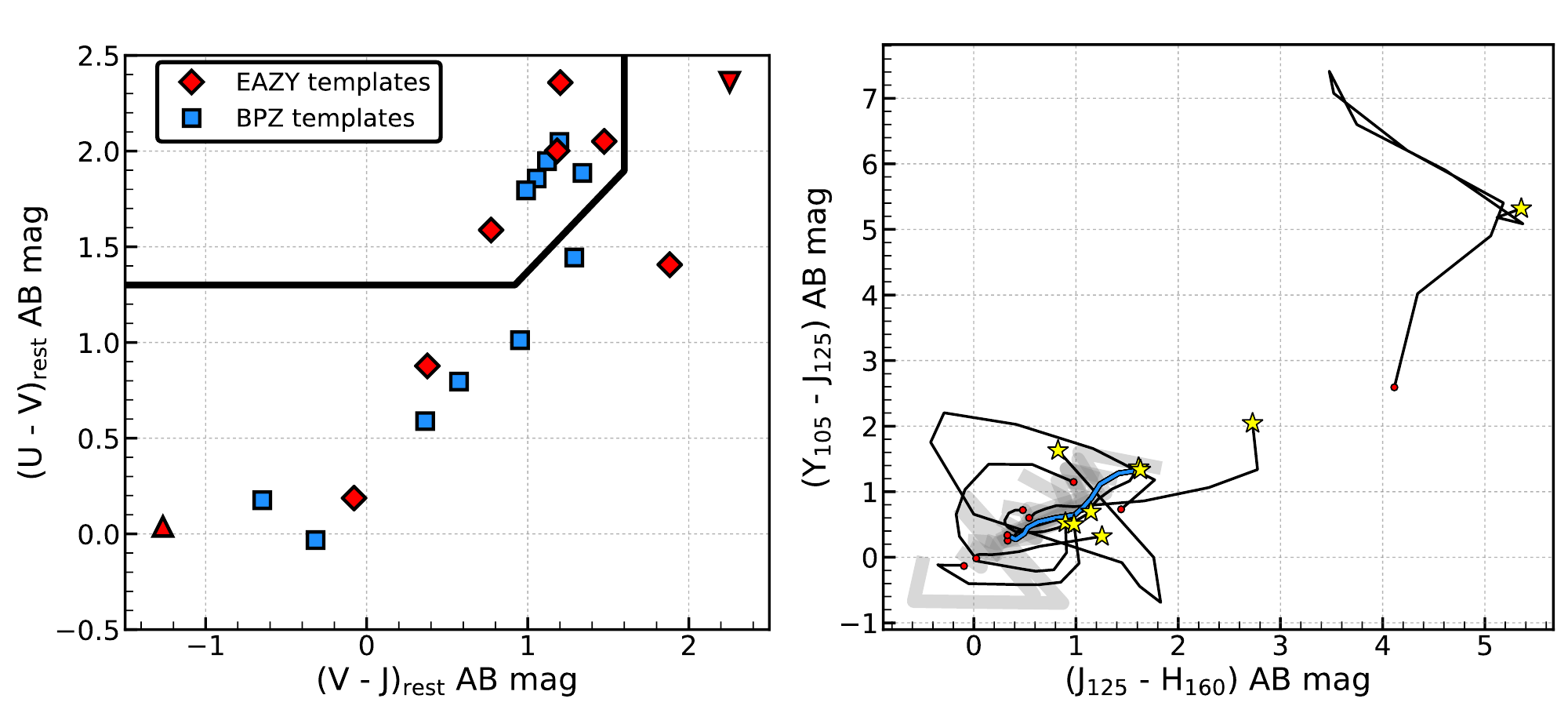}}
\caption{
Left: Rest-frame UVJ colors of the default SED templates used by {EAZY} (red diamonds)
and {BPZ} (blue squares). The downward triangle is a very dusty SED
template and the upward triangle is an extreme nebular emission line
template from {EAZY}. While both codes use linear combinations of these
templates to span a wide range of colors, the different assumed templates 
are the primary source for any photometric-redshift disagreement. 
Right: WFC3 YJH colors for the nine default {EAZY} SED templates. 
The gray thick lines follow the colors of each template with increasing redshift from $z=0.5$ to
the red circle at $z=2$. The black line continues to show the colors with increasing redshift until
the yellow star at $z=11$. The blue line shows the extreme nebular emission line template, which has very
degenerate colors at $z>5$ with those at $z<3$. The convergence of the $z$=10--11 yellow stars emphasizes the difficulty in distinguishing between galaxies at $z<2$ and $z>9$. 
}
\vspace{0.2 cm}      
\label{fig:RestFrame}
\end{figure*} 
\clearpage

\begin{sidewaystable}
\caption{$z\sim$ 6 Galaxy Candidates Behind 41 RELICS Clusters}\label{tab:z6}
\smallskip
\begin{tabular*}{\textwidth}{l @{\extracolsep{\fill}} *{11}{l}}
\toprule
Object ID$^{\rm a}$ & 
$\alpha_{J2000}$ & 
$\delta_{J2000}$ & 
$B_{435}$ & 
$V_{606}$ & 
$I_{814}$ & 
$Y_{105}$ & 
$J_{125}$ & 
$JH_{140}$ & 
$H_{160}$ & 
$z_{\rm EZ}$$^{\rm b}$ & 
$z_{\rm BPZ}$$^{\rm c}$ \\
\midrule
RXS0603+42-N-1888 & 90.793252 & 42.270382 & . . . & $26.31\!\pm\! 0.18$ & $23.89\!\pm\! 0.01$ & $22.99\!\pm\! 0.02$ & $22.75\!\pm\! 0.02$ & $22.75\!\pm\! 0.02$ & $22.78\!\pm\! 0.01$ & $5.4^{+0.2}_{-0.4}$ & $5.6^{+0.1}_{-0.1}$ \\
RXS0603+42-S-2085 & 90.864338 & 42.182232 & . . . & $27.38\!\pm\! 0.36$ & $25.01\!\pm\! 0.03$ & $23.61\!\pm\! 0.03$ & $23.19\!\pm\! 0.04$ & $23.21\!\pm\! 0.03$ & $23.21\!\pm\! 0.02$ & $5.9^{+0.3}_{-0.3}$ & $6.0^{+0.1}_{-0.1}$ \\
MACS0308+26-0904 & 47.222529 & 26.749704 & $26.54\!\pm\! 0.49$ & $>27.8$ & $25.12\!\pm\! 0.12$ & $23.30\!\pm\! 0.03$ & $23.20\!\pm\! 0.05$ & $23.37\!\pm\! 0.04$ & $23.36\!\pm\! 0.03$ & $6.4^{+0.2}_{-0.2}$ & $6.3^{+0.1}_{-0.1}$ \\
RXS0603+42-S-3486 & 90.853690 & 42.171170 & . . . & $27.10\!\pm\! 0.30$ & $24.58\!\pm\! 0.02$ & $23.78\!\pm\! 0.03$ & $23.40\!\pm\! 0.03$ & $23.47\!\pm\! 0.03$ & $23.41\!\pm\! 0.02$ & $5.4^{+0.3}_{-0.5}$ & $5.6^{+0.1}_{-0.1}$ \\
RXS0603+42-N-5737 & 90.788352 & 42.248042 & . . . & $26.65\!\pm\! 0.21$ & $24.67\!\pm\! 0.02$ & $23.69\!\pm\! 0.03$ & $23.55\!\pm\! 0.04$ & $23.48\!\pm\! 0.03$ & $23.48\!\pm\! 0.02$ & $1.0^{+4.8}_{-0.1}$ & $5.7^{+0.1}_{-0.1}$ \\
RXS0603+42-S-6059 & 90.863787 & 42.150244 & . . . & $26.79\!\pm\! 0.25$ & $24.77\!\pm\! 0.03$ & $23.79\!\pm\! 0.03$ & $23.53\!\pm\! 0.04$ & $23.53\!\pm\! 0.03$ & $23.54\!\pm\! 0.02$ & $1.0^{+4.9}_{-0.0}$ & $5.7^{+0.1}_{-0.1}$ \\
RXS0603+42-N-1050 & 90.786687 & 42.276517 & . . . & $26.72\!\pm\! 0.39$ & $25.49\!\pm\! 0.05$ & $23.88\!\pm\! 0.03$ & $23.65\!\pm\! 0.04$ & $23.58\!\pm\! 0.04$ & $23.57\!\pm\! 0.02$ & $6.1^{+0.3}_{-0.2}$ & $6.1^{+0.1}_{-0.1}$ \\
SMACS0723-73-1366 & 110.813866 & -73.469150 & $27.68\!\pm\! 0.67$ & $29.08\!\pm\! 1.02$ & $25.22\!\pm\! 0.07$ & $24.10\!\pm\! 0.03$ & $23.85\!\pm\! 0.04$ & $23.88\!\pm\! 0.04$ & $23.87\!\pm\! 0.03$ & $5.9^{+0.3}_{-0.3}$ & $5.8^{+0.1}_{-0.1}$ \\
RXC0600-20-0840 & 90.039804 & -20.136336 & $25.80\!\pm\! 0.22$ & $>28.2$ & $25.99\!\pm\! 0.16$ & $24.01\!\pm\! 0.04$ & $23.88\!\pm\! 0.07$ & $23.92\!\pm\! 0.06$ & $23.93\!\pm\! 0.04$ & $6.4^{+0.3}_{-0.2}$ & $6.3^{+0.2}_{-0.2}$ \\
PLCKG171-40-0130 & 48.225851 & 8.384411 & $>25.9$ & $>26.8$ & $25.15\!\pm\! 0.20$ & $24.22\!\pm\! 0.10$ & $23.93\!\pm\! 0.12$ & $24.07\!\pm\! 0.11$ & $23.93\!\pm\! 0.06$ & $5.8^{+0.4}_{-1.1}$ & $5.7^{+0.3}_{-5.1}$ \\
SMACS0723-73-0711 & 110.810518 & -73.452826 & $27.22\!\pm\! 0.59$ & $27.42\!\pm\! 0.39$ & $25.09\!\pm\! 0.07$ & $24.21\!\pm\! 0.04$ & $24.08\!\pm\! 0.06$ & $23.99\!\pm\! 0.05$ & $23.96\!\pm\! 0.03$ & $5.5^{+0.5}_{-0.8}$ & $5.6^{+0.1}_{-0.2}$ \\
PLCKG004-19-0222 & 289.282178 & -33.508191 & $28.51\!\pm\! 0.87$ & $28.28\!\pm\! 0.56$ & $25.38\!\pm\! 0.07$ & $24.33\!\pm\! 0.06$ & $24.00\!\pm\! 0.07$ & $24.20\!\pm\! 0.07$ & $24.13\!\pm\! 0.05$ & $5.8^{+0.3}_{-0.4}$ & $5.8^{+0.1}_{-0.1}$ \\
SPT0615-57-0261 & 93.955568 & -57.770519 & $>28.2$ & $28.64\!\pm\! 0.67$ & $25.50\!\pm\! 0.05$ & $24.66\!\pm\! 0.05$ & $24.56\!\pm\! 0.08$ & $24.38\!\pm\! 0.06$ & $24.21\!\pm\! 0.04$ & $5.7^{+0.4}_{-0.9}$ & $5.6^{+0.1}_{-4.8}$ \\
RXC2211-03-0196 & 332.945088 & -3.816933 & $>28.2$ & $>28.9$ & $26.13\!\pm\! 0.13$ & $24.43\!\pm\! 0.04$ & $24.18\!\pm\! 0.06$ & $24.36\!\pm\! 0.06$ & $24.23\!\pm\! 0.03$ & $6.2^{+0.3}_{-0.2}$ & $6.1^{+0.1}_{-0.1}$ \\
RXS0603+42-N-6150 & 90.790722 & 42.245871 & . . . & $27.17\!\pm\! 0.29$ & $25.42\!\pm\! 0.04$ & $24.53\!\pm\! 0.04$ & $24.25\!\pm\! 0.06$ & $24.32\!\pm\! 0.05$ & $24.31\!\pm\! 0.03$ & $1.0^{+4.9}_{-0.1}$ & $5.6^{+0.1}_{-0.1}$ \\
CL0152-13-1508 & 28.165607 & -13.968644 & $>28.1$ & . . . & . . . & $24.77\!\pm\! 0.07$ & $24.54\!\pm\! 0.04$ & $24.46\!\pm\! 0.08$ & $24.38\!\pm\! 0.03$ & $5.5^{+0.3}_{-0.2}$ & $5.6^{+0.1}_{-0.1}$ \\
MACS0553-33-1323 & 88.355157 & -33.726948 & $>29.1$ & $27.98\!\pm\! 0.32$ & $25.58\!\pm\! 0.04$ & $24.60\!\pm\! 0.05$ & $24.40\!\pm\! 0.06$ & $24.48\!\pm\! 0.06$ & $24.42\!\pm\! 0.04$ & $5.7^{+0.4}_{-0.4}$ & $5.7^{+0.1}_{-0.1}$ \\
RXC0142+44-0323 & 25.752926 & 44.642511 & $28.02\!\pm\! 0.58$ & $27.70\!\pm\! 0.32$ & $25.37\!\pm\! 0.06$ & $24.55\!\pm\! 0.04$ & $24.51\!\pm\! 0.07$ & $24.41\!\pm\! 0.05$ & $24.44\!\pm\! 0.04$ & $5.5^{+0.4}_{-0.5}$ & $5.5^{+0.3}_{-0.1}$ \\
Abell2537-0333 & 347.104166 & -2.188587 & $>27.4$ & $28.76\!\pm\! 1.24$ & $25.59\!\pm\! 0.16$ & $24.50\!\pm\! 0.08$ & $24.44\!\pm\! 0.14$ & $24.32\!\pm\! 0.11$ & $24.44\!\pm\! 0.08$ & $5.8^{+0.4}_{-0.5}$ & $5.8^{+0.3}_{-0.3}$ \\
PLCKG004-19-2691 & 289.265221 & -33.544507 & $27.60\!\pm\! 0.44$ & $> 0.0$ & $26.06\!\pm\! 0.11$ & $24.70\!\pm\! 0.08$ & $24.38\!\pm\! 0.09$ & $24.52\!\pm\! 0.09$ & $24.47\!\pm\! 0.05$ & $6.0^{+0.3}_{-0.4}$ & $6.0^{+0.1}_{-0.2}$ \\
PLCKG004-19-1483 & 289.287163 & -33.524637 & $29.15\!\pm\! 1.31$ & $27.64\!\pm\! 0.35$ & $25.05\!\pm\! 0.05$ & $24.37\!\pm\! 0.04$ & $24.27\!\pm\! 0.06$ & $24.17\!\pm\! 0.05$ & $24.47\!\pm\! 0.05$ & $5.5^{+0.4}_{-0.2}$ & $5.7^{+0.1}_{-0.3}$ \\
Abell2163-1204 & 243.918954 & -6.147594 & $>27.3$ & $27.97\!\pm\! 0.72$ & $25.05\!\pm\! 0.06$ & $24.43\!\pm\! 0.06$ & $24.50\!\pm\! 0.10$ & $24.38\!\pm\! 0.08$ & $24.49\!\pm\! 0.05$ & $5.6^{+0.4}_{-0.4}$ & $5.6^{+0.1}_{-0.4}$ \\
SMACS0723-73-1430 & 110.859546 & -73.472110 & $>27.9$ & $28.21\!\pm\! 0.57$ & $25.51\!\pm\! 0.09$ & $24.73\!\pm\! 0.05$ & $24.58\!\pm\! 0.08$ & $24.64\!\pm\! 0.07$ & $24.51\!\pm\! 0.04$ & $5.5^{+0.5}_{-0.6}$ & $5.5^{+0.2}_{-0.2}$ \\
RXS0603+42-N-1925 & 90.824322 & 42.270199 & . . . & $28.51\!\pm\! 0.75$ & $25.76\!\pm\! 0.06$ & $24.75\!\pm\! 0.05$ & $24.53\!\pm\! 0.07$ & $24.59\!\pm\! 0.06$ & $24.52\!\pm\! 0.04$ & $5.7^{+0.4}_{-0.4}$ & $5.7^{+0.1}_{-0.1}$ \\
RXS0603+42-N-4609 & 90.786605 & 42.253922 & . . . & $26.97\!\pm\! 0.22$ & $26.28\!\pm\! 0.08$ & $24.97\!\pm\! 0.06$ & $24.67\!\pm\! 0.07$ & $24.67\!\pm\! 0.07$ & $24.63\!\pm\! 0.04$ & $1.4^{+0.1}_{-0.2}$ & $6.0^{+0.1}_{-5.0}$ \\
RXS0603+42-N-7437 & 90.810315 & 42.239204 & . . . & $>28.7$ & $25.62\!\pm\! 0.04$ & $24.73\!\pm\! 0.04$ & $24.48\!\pm\! 0.06$ & $24.60\!\pm\! 0.05$ & $24.65\!\pm\! 0.04$ & $5.9^{+0.2}_{-0.3}$ & $5.6^{+0.3}_{-0.1}$ \\
RXS0603+42-N-4419 & 90.775132 & 42.255014 & . . . & $28.44\!\pm\! 0.57$ & $25.73\!\pm\! 0.04$ & $24.75\!\pm\! 0.06$ & $24.53\!\pm\! 0.09$ & $24.68\!\pm\! 0.08$ & $24.66\!\pm\! 0.05$ & $5.8^{+0.4}_{-0.4}$ & $5.6^{+0.3}_{-0.1}$ \\
Abell2537-0397 & 347.105646 & -2.188369 & $>28.4$ & $> 0.0$ & $25.43\!\pm\! 0.06$ & $24.72\!\pm\! 0.04$ & $24.46\!\pm\! 0.06$ & $24.80\!\pm\! 0.07$ & $24.67\!\pm\! 0.04$ & $5.7^{+0.4}_{-0.3}$ & $5.6^{+0.2}_{-0.3}$ \\
RXC1514-15-0550 & 228.758541 & -15.382887 & $>28.1$ & $31.41\!\pm\! 2.78$ & $26.20\!\pm\! 0.15$ & $24.98\!\pm\! 0.07$ & $24.76\!\pm\! 0.09$ & $24.70\!\pm\! 0.07$ & $24.74\!\pm\! 0.05$ & $5.9^{+0.4}_{-0.5}$ & $5.9^{+0.2}_{-0.2}$ \\
SPT0615-57-0325 & 93.978235 & -57.771646 & $29.09\!\pm\! 1.00$ & $28.94\!\pm\! 0.65$ & $26.33\!\pm\! 0.06$ & $25.00\!\pm\! 0.06$ & $24.75\!\pm\! 0.07$ & $24.78\!\pm\! 0.06$ & $24.78\!\pm\! 0.04$ & $6.0^{+0.3}_{-0.3}$ & $6.0^{+0.1}_{-0.1}$ \\
MACS0308+26-0438 & 47.238287 & 26.763436 & $25.93\!\pm\! 0.22$ & $>28.4$ & $26.75\!\pm\! 0.33$ & $24.77\!\pm\! 0.06$ & $24.64\!\pm\! 0.09$ & $24.74\!\pm\! 0.08$ & $24.78\!\pm\! 0.06$ & $6.4^{+0.5}_{-0.3}$ & $6.3^{+0.5}_{-0.3}$ \\
MACS0308+26-0249 & 47.237605 & 26.768654 & $25.95\!\pm\! 0.24$ & $27.25\!\pm\! 0.36$ & $25.45\!\pm\! 0.12$ & $24.71\!\pm\! 0.07$ & $24.75\!\pm\! 0.11$ & $24.78\!\pm\! 0.09$ & $24.78\!\pm\! 0.06$ & $1.0^{+0.3}_{-0.1}$ & $5.6^{+0.2}_{-0.5}$ \\
PLCKG209+10-0202 & 110.587210 & 7.420105 & $27.96\!\pm\! 0.59$ & $>28.9$ & $26.39\!\pm\! 0.17$ & $25.00\!\pm\! 0.06$ & $24.80\!\pm\! 0.09$ & $24.72\!\pm\! 0.07$ & $24.79\!\pm\! 0.05$ & $6.0^{+0.4}_{-0.4}$ & $6.0^{+0.2}_{-0.2}$ \\
Abell1758-1942 & 203.200110 & 50.518517 & $>28.7$ & $>29.1$ & $25.74\!\pm\! 0.05$ & $24.69\!\pm\! 0.05$ & $24.82\!\pm\! 0.09$ & $24.77\!\pm\! 0.07$ & $24.82\!\pm\! 0.05$ & $6.2^{+0.1}_{-0.4}$ & $6.0^{+0.1}_{-0.2}$ \\
RXC0600-20-1304 & 90.032239 & -20.152614 & $27.12\!\pm\! 0.30$ & $29.16\!\pm\! 0.84$ & $26.16\!\pm\! 0.10$ & $24.81\!\pm\! 0.06$ & $24.75\!\pm\! 0.09$ & $24.82\!\pm\! 0.09$ & $24.85\!\pm\! 0.06$ & $6.1^{+0.2}_{-0.3}$ & $6.1^{+0.1}_{-0.3}$ \\
Abell2163-1163 & 243.944229 & -6.146962 & $>27.7$ & $29.21\!\pm\! 1.23$ & $26.54\!\pm\! 0.15$ & $24.87\!\pm\! 0.06$ & $24.69\!\pm\! 0.09$ & $24.64\!\pm\! 0.07$ & $24.86\!\pm\! 0.06$ & $6.2^{+0.3}_{-0.2}$ & $6.2^{+0.2}_{-0.2}$ \\
Abell2163-1612 & 243.918938 & -6.155229 & $28.03\!\pm\! 0.89$ & $28.70\!\pm\! 0.86$ & $25.96\!\pm\! 0.08$ & $24.90\!\pm\! 0.07$ & $24.84\!\pm\! 0.10$ & $24.88\!\pm\! 0.09$ & $24.94\!\pm\! 0.06$ & $6.0^{+0.3}_{-0.4}$ & $6.0^{+0.1}_{-0.4}$ \\
PLCKG287+32-3078 & 177.708907 & -28.097272 & $>27.9$ & $27.39\!\pm\! 0.40$ & $27.44\!\pm\! 0.40$ & $25.41\!\pm\! 0.13$ & $24.90\!\pm\! 0.14$ & $25.16\!\pm\! 0.15$ & $24.96\!\pm\! 0.08$ & $1.4^{+6.2}_{-0.2}$ & $6.5^{+0.9}_{-0.5}$ \\
Abells295-0250 & 41.374456 & -53.030668 & $28.13\!\pm\! 0.69$ & $>28.8$ & $27.05\!\pm\! 0.20$ & $25.10\!\pm\! 0.07$ & $24.93\!\pm\! 0.11$ & $25.09\!\pm\! 0.11$ & $25.04\!\pm\! 0.07$ & $6.3^{+0.3}_{-0.2}$ & $6.3^{+0.3}_{-0.2}$ \\
RXS0603+42-S-5318 & 90.835657 & 42.156162 & . . . & $28.06\!\pm\! 0.47$ & $26.87\!\pm\! 0.12$ & $25.47\!\pm\! 0.09$ & $25.04\!\pm\! 0.10$ & $25.30\!\pm\! 0.11$ & $25.12\!\pm\! 0.06$ & $1.4^{+5.0}_{-0.2}$ & $6.0^{+0.1}_{-4.9}$ \\
RXC2211-03-0547 & 332.929227 & -3.827773 & $>28.1$ & $>28.7$ & $25.50\!\pm\! 0.08$ & $25.20\!\pm\! 0.09$ & $25.09\!\pm\! 0.14$ & $25.18\!\pm\! 0.13$ & $25.20\!\pm\! 0.09$ & $5.7^{+0.2}_{-0.5}$ & $5.5^{+0.1}_{-0.6}$ \\
\bottomrule
\end{tabular*}
\begin{tablenotes}
\footnotesize
\item {\bf Notes:} The full tables of the $z$=6 sample, including all ancillary HST data, will be made available in the online journal version. The brightest (in H160) 40 candidates are shown here as an example of the format. All magnitudes are given as observed (lensed) isophotal AB magnitudes. \\
$^{\rm a}$ The online tables will use the full cluster name and ID as used by the released RELICS photometric catalogs. \\
$^{\rm b}$ The EAZY photometric redshifts assuming a flat prior in magnitude, along with their 2-$\sigma$ uncertainty. Cases where the uncertainty reaches $\Delta z>2$ are due to a secondary peak in probability at lower redshift. \\
$^{\rm c}$ The BPZ photometric redshifts assuming a small prior in magnitude, along with their 2-$\sigma$ uncertainty. Throughout this work, the redshift we assume for each candidate is the average of the BPZ and EAZY estimates unless they differ by $\Delta z>1$ in which case we adopt the higher redshift solution. 
\end{tablenotes}

\end{sidewaystable}
\global \pdfpageattr\expandafter{\the\pdfpageattr/Rotate 270}
\clearpage

\begin{sidewaystable}

\caption{$z\sim$ 7 Galaxy Candidates Behind 41 RELICS Clusters}\label{tab:z7}
\smallskip
\begin{tabular*}{\textwidth}{l @{\extracolsep{\fill}} *{11}{l}}
\toprule
Object ID & 
$\alpha_{J2000}$ & 
$\delta_{J2000}$ & 
$B_{435}$ & 
$V_{606}$ & 
$I_{814}$ & 
$Y_{105}$ & 
$J_{125}$ & 
$JH_{140}$ & 
$H_{160}$ & 
$z_{\rm EZ}$ & 
$z_{\rm BPZ}$ \\
\midrule
RXC0600-20-0178 & 90.027105 & -20.120249 & $30.24\!\pm\! 1.99$ & $27.99\!\pm\! 0.33$ & $27.55\!\pm\! 0.29$ & $25.01\!\pm\! 0.06$ & $24.67\!\pm\! 0.08$ & $24.55\!\pm\! 0.06$ & $24.42\!\pm\! 0.04$ & $6.5^{+0.3}_{-0.4}$ & $6.7^{+0.5}_{-0.1}$ \\
MS1008-12-0326 & 152.630990 & -12.653099 & $>28.1$ & $28.40\!\pm\! 0.70$ & . . . & $25.26\!\pm\! 0.09$ & $24.83\!\pm\! 0.10$ & $24.68\!\pm\! 0.07$ & $24.55\!\pm\! 0.05$ & $7.2^{+0.7}_{-0.6}$ & $7.2^{+0.3}_{-0.3}$ \\
PLCKG287+32-0698 & 177.704968 & -28.070710 & $28.95\!\pm\! 1.21$ & $>28.5$ & $28.59\!\pm\! 0.78$ & $25.11\!\pm\! 0.08$ & $24.95\!\pm\! 0.12$ & $24.98\!\pm\! 0.10$ & $25.00\!\pm\! 0.07$ & $6.9^{+0.7}_{-0.3}$ & $6.7^{+0.5}_{-0.3}$ \\
PLCKG287+32-0777 & 177.719902 & -28.071524 & $28.73\!\pm\! 1.00$ & $28.47\!\pm\! 0.64$ & $>28.7$ & $25.38\!\pm\! 0.09$ & $25.04\!\pm\! 0.11$ & $25.09\!\pm\! 0.10$ & $25.16\!\pm\! 0.07$ & $7.5^{+0.2}_{-0.7}$ & $7.2^{+0.2}_{-0.6}$ \\
PLCKG287+32-2235 & 177.716360 & -28.087833 & $>28.1$ & $28.84\!\pm\! 0.90$ & $>28.6$ & $25.33\!\pm\! 0.09$ & $25.14\!\pm\! 0.13$ & $25.18\!\pm\! 0.12$ & $25.17\!\pm\! 0.07$ & $7.3^{+0.3}_{-0.5}$ & $7.0^{+0.3}_{-0.5}$ \\
Abell3192-1076 & 59.711133 & -29.946217 & $29.05\!\pm\! 0.70$ & $32.74\!\pm\! 3.39$ & $29.90\!\pm\! 1.17$ & $25.59\!\pm\! 0.09$ & $25.34\!\pm\! 0.14$ & $25.40\!\pm\! 0.12$ & $25.32\!\pm\! 0.08$ & $7.1^{+0.6}_{-0.4}$ & $6.8^{+0.5}_{-0.2}$ \\
Abell1300-0226 & 172.983602 & -19.915016 & $27.77\!\pm\! 0.39$ & $30.15\!\pm\! 1.28$ & $29.06\!\pm\! 0.90$ & $25.97\!\pm\! 0.11$ & $25.58\!\pm\! 0.13$ & $25.92\!\pm\! 0.15$ & $25.62\!\pm\! 0.08$ & $7.3^{+0.4}_{-1.0}$ & $6.7^{+0.7}_{-0.4}$ \\
RXC0911+17-0419 & 137.787292 & 17.781433 & . . . & . . . & $28.87\!\pm\! 0.50$ & $25.36\!\pm\! 0.09$ & $25.65\!\pm\! 0.18$ & $25.25\!\pm\! 0.11$ & $25.64\!\pm\! 0.10$ & $6.9^{+0.3}_{-0.2}$ & $6.7^{+0.2}_{-0.1}$ \\
Abell1300-0534 & 172.985370 & -19.924202 & $27.81\!\pm\! 0.42$ & $>29.2$ & $28.97\!\pm\! 0.89$ & $26.00\!\pm\! 0.12$ & $25.84\!\pm\! 0.18$ & $25.82\!\pm\! 0.15$ & $25.94\!\pm\! 0.11$ & $6.8^{+0.8}_{-0.5}$ & $6.6^{+0.6}_{-0.4}$ \\
Abell3192-0033 & 59.719401 & -29.911883 & $>29.4$ & $29.09\!\pm\! 0.50$ & $28.84\!\pm\! 0.48$ & $26.07\!\pm\! 0.11$ & $26.44\!\pm\! 0.28$ & $26.05\!\pm\! 0.17$ & $26.07\!\pm\! 0.12$ & $6.8^{+0.4}_{-0.5}$ & $6.6^{+0.4}_{-0.3}$ \\
PLCKG287+32-0792 & 177.700898 & -28.071636 & $>28.8$ & $29.47\!\pm\! 0.91$ & $>29.2$ & $26.79\!\pm\! 0.19$ & $26.71\!\pm\! 0.30$ & $26.37\!\pm\! 0.19$ & $26.24\!\pm\! 0.11$ & $7.3^{+0.6}_{-1.1}$ & $1.6^{+6.2}_{-0.3}$ \\
ACT0102-49-2551 & 15.725680 & -49.232616 & $29.16\!\pm\! 0.86$ & $28.97\!\pm\! 0.56$ & $29.66\!\pm\! 1.18$ & $25.98\!\pm\! 0.10$ & $25.79\!\pm\! 0.15$ & $25.92\!\pm\! 0.14$ & $26.25\!\pm\! 0.12$ & $7.0^{+0.3}_{-0.4}$ & $6.7^{+0.3}_{-0.2}$ \\
ACT0102-49-2391 & 15.723132 & -49.239372 & $27.90\!\pm\! 0.37$ & $>29.3$ & $>28.8$ & $26.39\!\pm\! 0.15$ & $25.99\!\pm\! 0.19$ & $26.29\!\pm\! 0.21$ & $26.26\!\pm\! 0.14$ & $7.0^{+0.7}_{-0.5}$ & $6.7^{+0.5}_{-0.4}$ \\
MS1008-12-0312 & 152.647497 & -12.652450 & $>28.8$ & $29.15\!\pm\! 0.76$ & . . . & $26.50\!\pm\! 0.15$ & $26.40\!\pm\! 0.22$ & $26.60\!\pm\! 0.22$ & $26.31\!\pm\! 0.12$ & $7.1^{+0.5}_{-2.2}$ & $6.8^{+0.5}_{-5.5}$ \\
PLCKG287+32-2252 & 177.703898 & -28.084229 & $>28.7$ & $>29.2$ & $28.57\!\pm\! 0.46$ & $26.73\!\pm\! 0.18$ & $26.85\!\pm\! 0.32$ & $26.90\!\pm\! 0.28$ & $26.40\!\pm\! 0.13$ & $6.5^{+0.7}_{-4.7}$ & $1.2^{+4.8}_{-0.4}$ \\
MS1008-12-0313 & 152.647304 & -12.652520 & $28.13\!\pm\! 0.49$ & $>29.1$ & . . . & $26.49\!\pm\! 0.16$ & $26.52\!\pm\! 0.26$ & $26.37\!\pm\! 0.19$ & $26.46\!\pm\! 0.14$ & $6.8^{+0.7}_{-0.7}$ & $6.4^{+0.6}_{-5.3}$ \\
Abell1763-1500 & 203.831009 & 40.988866 & $>28.9$ & $30.76\!\pm\! 1.66$ & $29.18\!\pm\! 0.85$ & $26.55\!\pm\! 0.14$ & $26.84\!\pm\! 0.13$ & $26.35\!\pm\! 0.22$ & $26.50\!\pm\! 0.09$ & $6.7^{+0.6}_{-1.0}$ & $6.4^{+0.5}_{-5.0}$ \\
SPT0254-58-0476 & 43.593763 & -58.949693 & $>29.1$ & $28.78\!\pm\! 0.42$ & $28.82\!\pm\! 0.56$ & $26.52\!\pm\! 0.15$ & $26.84\!\pm\! 0.33$ & $26.49\!\pm\! 0.21$ & $26.50\!\pm\! 0.14$ & $6.7^{+0.6}_{-5.5}$ & $6.4^{+0.6}_{-5.3}$ \\
RXC0032+18-0571 & 8.052543 & 18.131884 & $28.81\!\pm\! 0.88$ & $31.32\!\pm\! 2.32$ & $28.74\!\pm\! 0.67$ & $26.48\!\pm\! 0.16$ & $26.42\!\pm\! 0.26$ & $26.82\!\pm\! 0.29$ & $26.50\!\pm\! 0.16$ & $6.7^{+0.7}_{-1.5}$ & $6.4^{+0.7}_{-0.8}$ \\
SMACS0723-73-1317 & 110.785854 & -73.467361 & $28.78\!\pm\! 0.71$ & $>29.6$ & $>29.2$ & $26.24\!\pm\! 0.14$ & $26.43\!\pm\! 0.23$ & $26.56\!\pm\! 0.23$ & $26.53\!\pm\! 0.15$ & $7.0^{+0.5}_{-0.4}$ & $6.7^{+0.4}_{-0.4}$ \\
Abell520-1342 & 73.531755 & 2.903624 & $28.88\!\pm\! 0.70$ & $28.88\!\pm\! 0.56$ & $>29.2$ & $26.86\!\pm\! 0.20$ & $26.53\!\pm\! 0.26$ & $26.36\!\pm\! 0.19$ & $26.61\!\pm\! 0.16$ & $7.6^{+0.4}_{-1.1}$ & $6.9^{+0.3}_{-5.9}$ \\
CL0152-13-0410 & 28.182402 & -13.946921 & $>28.9$ & . . . & . . . & $26.46\!\pm\! 0.15$ & $26.87\!\pm\! 0.14$ & $26.58\!\pm\! 0.24$ & $26.66\!\pm\! 0.12$ & $6.7^{+0.2}_{-0.4}$ & $6.4^{+0.2}_{-0.3}$ \\
RXC0142+44-1088 & 25.714213 & 44.623339 & $30.41\!\pm\! 1.54$ & $>29.7$ & $>29.4$ & $27.23\!\pm\! 0.22$ & $27.00\!\pm\! 0.28$ & $27.30\!\pm\! 0.30$ & $26.68\!\pm\! 0.13$ & $7.0^{+1.0}_{-1.9}$ & $1.3^{+6.2}_{-0.3}$ \\
PLCKG287+32-2457 & 177.706999 & -28.090180 & $28.81\!\pm\! 0.78$ & $>28.9$ & $29.68\!\pm\! 1.17$ & $27.15\!\pm\! 0.24$ & $27.46\!\pm\! 0.54$ & $27.63\!\pm\! 0.48$ & $26.69\!\pm\! 0.17$ & $6.7^{+1.0}_{-5.6}$ & $1.4^{+5.7}_{-0.7}$ \\
CL0152-13-0259 & 28.182517 & -13.971709 & $>28.8$ & . . . & . . . & $26.40\!\pm\! 0.16$ & $26.38\!\pm\! 0.10$ & $26.36\!\pm\! 0.23$ & $26.70\!\pm\! 0.14$ & $7.0^{+0.5}_{-0.5}$ & $6.7^{+0.4}_{-0.4}$ \\
MS1008-12-0227 & 152.648835 & -12.649399 & $>28.9$ & $>29.1$ & . . . & $27.10\!\pm\! 0.24$ & $27.59\!\pm\! 0.55$ & $26.74\!\pm\! 0.24$ & $26.72\!\pm\! 0.17$ & $7.2^{+0.7}_{-2.9}$ & $1.3^{+6.1}_{-0.4}$ \\
PLCKG004-19-2704 & 289.272136 & -33.544955 & $>28.7$ & $29.28\!\pm\! 0.83$ & $>28.9$ & $26.91\!\pm\! 0.20$ & $26.68\!\pm\! 0.28$ & $27.13\!\pm\! 0.32$ & $26.72\!\pm\! 0.17$ & $7.0^{+0.9}_{-0.7}$ & $6.7^{+0.7}_{-5.6}$ \\
PLCKG004-19-1983 & 289.261526 & -33.530310 & $28.28\!\pm\! 0.51$ & $>29.4$ & $28.87\!\pm\! 0.65$ & $26.55\!\pm\! 0.14$ & $26.71\!\pm\! 0.26$ & $26.72\!\pm\! 0.22$ & $26.75\!\pm\! 0.15$ & $6.7^{+0.6}_{-0.7}$ & $6.4^{+0.5}_{-0.5}$ \\
Abell2163-1715 & 243.931878 & -6.162199 & $>28.3$ & $30.59\!\pm\! 1.85$ & $>29.2$ & $26.90\!\pm\! 0.21$ & $26.78\!\pm\! 0.30$ & $26.97\!\pm\! 0.29$ & $26.77\!\pm\! 0.17$ & $7.0^{+0.8}_{-1.5}$ & $6.7^{+0.7}_{-0.8}$ \\
SMACS0723-73-0599 & 110.840288 & -73.449995 & $>28.5$ & $>29.2$ & $29.64\!\pm\! 1.22$ & $27.17\!\pm\! 0.23$ & $27.30\!\pm\! 0.42$ & $26.77\!\pm\! 0.23$ & $26.82\!\pm\! 0.16$ & $6.5^{+1.3}_{-5.6}$ & $1.3^{+6.2}_{-0.4}$ \\
PLCKG287+32-0221 & 177.714041 & -28.064780 & $>28.7$ & $>29.1$ & $29.49\!\pm\! 1.00$ & $26.75\!\pm\! 0.28$ & $26.72\!\pm\! 0.50$ & $26.79\!\pm\! 0.40$ & $26.85\!\pm\! 0.31$ & $6.8^{+1.2}_{-5.3}$ & $6.4^{+1.0}_{-0.6}$ \\
MACS0257-23-0787 & 44.303042 & -23.441137 & $>29.3$ & . . . & $>30.1$ & $27.49\!\pm\! 0.27$ & $27.05\!\pm\! 0.32$ & $27.31\!\pm\! 0.33$ & $26.94\!\pm\! 0.17$ & $7.5^{+0.9}_{-1.2}$ & $6.9^{+0.9}_{-5.5}$ \\
SPT0615-57-0548 & 93.973142 & -57.777418 & $28.72\!\pm\! 0.60$ & $>29.7$ & $29.70\!\pm\! 0.58$ & $27.09\!\pm\! 0.20$ & $26.94\!\pm\! 0.29$ & $26.70\!\pm\! 0.21$ & $26.95\!\pm\! 0.17$ & $6.6^{+1.2}_{-0.8}$ & $6.6^{+0.8}_{-0.4}$ \\
PLCKG287+32-1887 & 177.691357 & -28.083799 & $27.78\!\pm\! 0.36$ & $>29.3$ & $>29.3$ & $27.42\!\pm\! 0.30$ & $27.26\!\pm\! 0.42$ & $27.98\!\pm\! 0.62$ & $27.00\!\pm\! 0.20$ & $7.0^{+1.4}_{-5.7}$ & $1.3^{+6.0}_{-0.4}$ \\
PLCKG004-19-1156 & 289.272414 & -33.520949 & $30.16\!\pm\! 1.62$ & $30.32\!\pm\! 1.41$ & $>29.0$ & $27.21\!\pm\! 0.25$ & $26.93\!\pm\! 0.31$ & $27.05\!\pm\! 0.29$ & $27.01\!\pm\! 0.20$ & $7.4^{+0.7}_{-5.8}$ & $6.7^{+0.9}_{-5.6}$ \\
MACS0553-33-0463 & 88.335048 & -33.703438 & $>29.8$ & $29.65\!\pm\! 0.67$ & $29.78\!\pm\! 0.71$ & $27.36\!\pm\! 0.25$ & $27.34\!\pm\! 0.39$ & $27.32\!\pm\! 0.32$ & $27.15\!\pm\! 0.19$ & $6.6^{+1.2}_{-5.7}$ & $6.5^{+0.9}_{-5.4}$ \\
MACS0553-33-0865 & 88.327049 & -33.713029 & $>29.8$ & $29.05\!\pm\! 0.43$ & $30.30\!\pm\! 0.98$ & $26.96\!\pm\! 0.19$ & $27.10\!\pm\! 0.33$ & $27.54\!\pm\! 0.39$ & $27.16\!\pm\! 0.20$ & $6.8^{+0.7}_{-0.7}$ & $6.6^{+0.6}_{-0.6}$ \\
Abell1763-1564 & 203.830702 & 40.987242 & $>29.0$ & $31.92\!\pm\! 2.48$ & $>29.2$ & $27.05\!\pm\! 0.19$ & $27.01\!\pm\! 0.13$ & $27.16\!\pm\! 0.37$ & $27.18\!\pm\! 0.13$ & $7.1^{+0.7}_{-0.7}$ & $6.6^{+0.7}_{-0.6}$ \\
Abell1763-0817 & 203.807526 & 41.002688 & $>29.2$ & $29.57\!\pm\! 0.68$ & $31.39\!\pm\! 2.26$ & $27.49\!\pm\! 0.25$ & $26.99\!\pm\! 0.11$ & $27.15\!\pm\! 0.35$ & $27.19\!\pm\! 0.12$ & $7.7^{+0.6}_{-2.4}$ & $7.2^{+0.6}_{-1.3}$ \\
PLCKG287+32-2199 & 177.717592 & -28.087374 & $>29.1$ & $29.23\!\pm\! 0.62$ & $>29.6$ & $27.24\!\pm\! 0.21$ & $27.84\!\pm\! 0.54$ & $27.55\!\pm\! 0.38$ & $27.21\!\pm\! 0.19$ & $6.9^{+0.8}_{-5.4}$ & $2.5^{+1.1}_{-1.6}$ \\
Abell3192-0728 & 59.734516 & -29.933640 & $>30.1$ & $>30.2$ & $32.49\!\pm\! 2.54$ & $27.16\!\pm\! 0.18$ & $27.15\!\pm\! 0.30$ & $27.93\!\pm\! 0.47$ & $27.27\!\pm\! 0.19$ & $7.0^{+0.7}_{-0.5}$ & $6.7^{+0.6}_{-0.4}$ \\
\bottomrule
\end{tabular*}
\begin{tablenotes}
\footnotesize
\item {\bf Notes:} The full table of the $z$=7 sample, including all ancillary HST data, will be made available in the online journal version. The brightest (in H160) 40 candidates are shown here as an example of the format. All magnitudes are given as observed (lensed) isophotal AB magnitudes.
\end{tablenotes}
\end{sidewaystable}
\global \pdfpageattr\expandafter{\the\pdfpageattr/Rotate 90}
\clearpage

\begin{sidewaystable}

\caption{$z\sim$ 8 Galaxy Candidates Behind 41 RELICS Clusters}\label{tab:z8}
\smallskip
\begin{tabular*}{\textwidth}{l @{\extracolsep{\fill}} *{11}{l}}
\toprule
Object ID & 
$\alpha_{J2000}$ & 
$\delta_{J2000}$ & 
$B_{435}$ & 
$V_{606}$ & 
$I_{814}$ & 
$Y_{105}$ & 
$J_{125}$ & 
$JH_{140}$ & 
$H_{160}$ & 
$z_{\rm EZ}$ & 
$z_{\rm BPZ}$ \\
\midrule
PLCKG287+32-2013 & 177.687797 & -28.076086 & $28.13\!\pm\! 0.78$ & $>28.5$ & $29.87\!\pm\! 1.64$ & $25.61\!\pm\! 0.19$ & $24.97\!\pm\! 0.19$ & $24.84\!\pm\! 0.13$ & $24.88\!\pm\! 0.09$ & $7.6^{+0.7}_{-1.3}$ & $7.5^{+0.3}_{-0.8}$ \\
Abell1763-1434 & 203.833374 & 40.990179 & $>28.6$ & $>29.1$ & $>28.8$ & $27.70\!\pm\! 0.43$ & $26.40\!\pm\! 0.10$ & $26.13\!\pm\! 0.22$ & $26.17\!\pm\! 0.08$ & $8.4^{+1.2}_{-1.8}$ & $8.2^{+0.4}_{-6.9}$ \\
Abells295-0568 & 41.401024 & -53.040518 & $>28.9$ & $>29.5$ & $31.78\!\pm\! 2.48$ & $27.07\!\pm\! 0.22$ & $26.02\!\pm\! 0.17$ & $26.38\!\pm\! 0.19$ & $26.35\!\pm\! 0.12$ & $8.1^{+0.3}_{-1.7}$ & $7.7^{+0.4}_{-0.9}$ \\
RXC0911+17-0143 & 137.793971 & 17.789752 & . . . & . . . & $31.41\!\pm\! 1.85$ & $27.18\!\pm\! 0.24$ & $26.12\!\pm\! 0.17$ & $26.61\!\pm\! 0.21$ & $26.45\!\pm\! 0.13$ & $8.1^{+0.3}_{-1.6}$ & $7.7^{+0.4}_{-0.9}$ \\
PLCKG287+32-2032 & 177.722594 & -28.085070 & $28.33\!\pm\! 0.55$ & $28.00\!\pm\! 0.31$ & $>29.2$ & $27.81\!\pm\! 0.41$ & $26.59\!\pm\! 0.26$ & $26.87\!\pm\! 0.28$ & $26.74\!\pm\! 0.17$ & $8.2^{+0.5}_{-6.8}$ & $1.6^{+6.0}_{-0.5}$ \\
MACS0553-33-0219 & 88.354035 & -33.697948 & $30.36\!\pm\! 0.98$ & $>30.0$ & $29.78\!\pm\! 0.63$ & $28.11\!\pm\! 0.41$ & $26.75\!\pm\! 0.23$ & $26.85\!\pm\! 0.21$ & $27.19\!\pm\! 0.18$ & $8.3^{+0.5}_{-6.9}$ & $7.9^{+0.4}_{-6.6}$ \\
SPT0615-57-1048 & 93.969721 & -57.789586 & $>29.6$ & $33.58\!\pm\! 3.51$ & $30.97\!\pm\! 1.30$ & $28.15\!\pm\! 0.36$ & $27.36\!\pm\! 0.31$ & $28.06\!\pm\! 0.46$ & $27.73\!\pm\! 0.24$ & $7.7^{+0.7}_{-6.7}$ & $6.7^{+1.3}_{-5.5}$ \\
Abell1763-0460 & 203.824976 & 41.009117 & $28.74\!\pm\! 0.46$ & $>30.0$ & $29.50\!\pm\! 0.74$ & $29.94\!\pm\! 1.16$ & $27.58\!\pm\! 0.15$ & $27.26\!\pm\! 0.29$ & $27.78\!\pm\! 0.16$ & $8.5^{+1.2}_{-7.1}$ & $8.1^{+0.4}_{-7.2}$\\
\bottomrule
\end{tabular*}
\begin{tablenotes}
\footnotesize
\item {\bf Notes:} The full table of the $z$=8 sample, including all ancillary HST data, will be made available in the online journal version. All eight candidates are shown here as an example of the format. All magnitudes are given as observed (lensed) isophotal AB magnitudes.
\end{tablenotes}

\end{sidewaystable}
\global \pdfpageattr\expandafter{\the\pdfpageattr/Rotate 270}
\clearpage

\clearpage

\end{document}